\documentclass[a4paper,12pt]{article}
\DeclareMathSizes{12}{12.5}{10}{10}
\usepackage[left=2.3cm,bottom=3cm,right=2.3cm,top=3cm]{geometry}
\usepackage{overpic,youngtab}
\usepackage{subfigure,yfonts}
\usepackage[utf8]{inputenc}
\usepackage[latin,english]{babel}
\usepackage{amsmath}
\usepackage{verbatim}
\usepackage{amssymb}
\usepackage{epsfig}
\usepackage{epstopdf}
\usepackage{scalerel}
\usepackage{float}
\usepackage{fancybox}
\usepackage{color}
\usepackage[colorlinks=true,citecolor=blue,,linktocpage=true,linkcolor=blue,urlcolor=black]{hyperref}
\usepackage{mathabx}

\definecolor{ogreen}{rgb}{0,0.7,0}

\DeclareGraphicsExtensions{.pdf,.png,.jpg,.gif,.jpeg}
\def\be{\begin{equation}}
\def\ee{\end{equation}}
\def\bea#1\eea{\begin{align}#1\end{align}}
\def\pd{\partial}
\def\a{\alpha}
\def\b{\beta}
\def\g{\gamma}
\def\d{\delta}
\def\e{\epsilon}

\def\m{\mu}
\def\n{\nu}
\def\t{\tau}

\def\l{\lambda}

\def\r{\rho}

\def\bR{\widebar{R}}

\def\bn{\widebar{\nabla}}
\def\bR{\widebar{R}}

\def\s{\sigma}
\def\e{\epsilon}
\def\t{\tau}
\def\bi{\begin{itemize}}
\def\ei{\end{itemize}}
\def\bg{\widebar{g}}
\def\bG{\widebar{\Gamma}}
\def\cO{\mathcal{O}}
\def\cR{\mathcal{R}}
\def\cL{\mathcal{L}}

\def\mup{\m^{\prime}}

\def\xp{x^{\prime}}

\def\scaleint#1{\vcenter{\hbox{\scaleto[3ex]{\displaystyle\int}{#1}}}}
\usepackage{authblk}
\usepackage{setspace}

\newcommand{\email}[1]{\href{mailto:#1}{\tt #1}}

\title{\centering\bf 
	Quadratic gravity in first order formalism}

\begin{document}


\vspace*{-1cm}
{\flushleft
	{{FTUAM-17-3}}
	\hfill{{IFT-UAM/CSIC-17-015}}}
\vskip 1.5cm
\begin{center}
	{\LARGE\bf Quadratic gravity in first order formalism}\\[3mm]
	\vskip .3cm
	
\end{center}
\vskip 0.5  cm
\begin{center}
	{\Large Enrique Alvarez}$^{~a}$, {\Large Jesus Anero}$^{~a}$
{{\large and} \Large Sergio Gonzalez-Martin}$^{~a}$
\\
\vskip .7cm
{
	$^{a}$Departamento de F\'isica Te\'orica and Instituto de F\'{\i}sica Te\'orica (IFT-UAM/CSIC),\\
	Universidad Aut\'onoma de Madrid, Cantoblanco, 28049, Madrid, Spain\\
	\vskip .1cm

	\vskip .5cm
	\begin{minipage}[l]{.9\textwidth}
		\begin{center} 
			\textit{E-mail:} 
			\email{enrique.alvarez@uam.es},
			\email{jesusanero@gmail.com},
			\email{sergio.gonzalez.martin@uam.es},
		\end{center}
	\end{minipage}
}
\end{center}
\thispagestyle{empty}

\begin{abstract}
	\noindent
	We consider the most general action for gravity which is quadratic in curvature. In this case first order and second order formalisms are not equivalent. This framework is a good candidate for a unitary and renormalizable theory of the gravitational field; in particular,  there are no propagators falling down faster than $\tfrac{1}{p^2}$. The drawback is of course that the parameter space of the theory is too big, so that in many cases will be far away from a theory of gravity alone. In order to analyze this issue, the interaction between external sources was examined in some detail. We find that  this interaction is conveyed mainly by propagation of  the three-index connection field. At any rate the theory as it stands is in the conformal invariant phase; only when Weyl invariance is broken through the coupling to matter can an Einstein-Hilbert term (and its corresponding Planck mass scale) be generated by quantum corrections.
\end{abstract}
\newpage
\tableofcontents

\newpage
\section{Introduction.}

It is well-known that general relativity  is not renormalizable
(cf.\cite{Alvarez} and references therein for a general review).
However, quadratic (in curvature) theories are renormalizable,
albeit not unitary \cite{Stelle} -at least in the standard second
order formalism- although they have been widely studied over the years \cite{Zee,Salvio}. When considering the Palatini version of the
Einstein-Hilbert lagrangian the connection and the metric are
treated as independent variables and the Levi-Civita connection
appears only when the equations of motion are used.
\par
It is however the case that when more general quadratic in curvature metric-affine actions are
considered  in first order formalism the deterministic relationship between the affine
connection and the Levi-Civita one is lost, even on shell. That
is, the equations of motion do not force the connection to be the
Levi-Civita one.
\par
This is quite interesting because it looks as if  we could  have all the goods of  quadratic lagrangians \cite{Stelle} (mainly renormalizability) without conflicting with  K\"allen-Lehmann's spectral theorem.

\par
The purpose of the present paper is to explore the most general
first order Weyl invariant quadratic lagrangian. By considering all possible monomials of a given symmetry the system is closed under renormalization in background field gauge. This work is a
continuation of \cite{AlvarezGM}(cf. also \cite{Donoghue}), whose
conventions we follow here. 
\par
A general issue when considering first order versus  second order theories is that in general the manifold of solutions in the first order treatment is too big. This means in our case that in many situations we are not dealing with a theory of gravity. One of our aims in this paper is to analyze the properties that physical sources need to have in order to reproduce a proper gravitational potential energy between static energy-momentum sources.

\par
Let us now summarize our general
framework.
\par
Let us start with some general remarks. An orthonormalized coframe will be characterized by n differential
forms \be
e^a\equiv e^a_\m dx^\m
\ee $a=1\ldots n$ are tangent (Lorentz) indices, and $\m,\n\ldots$
are spacetime (Einstein) indices. They obey \be
\eta_{ab}~e^a_\m(x)~ e^b_\n(x) = g_{\m\n}(x) \ee (where
$\eta_{ab}$ is the flat metric). Spacetime tensors are observed in
the frame as spacetime scalars, id est, \be V^a(x)\equiv
e^a_\m(x)~ V^\m(x) \ee
\par
The Lorentz (usually called spin) connection is defined by
demanding local Lorentz invariance of  derivatives of such scalars
as \be \nabla_\m V^b\equiv \pd_\m V^b+\omega_\m\,^b\,_c V^c \ee
Physical consistency demands that the Lorentz and Einstein
connections are equivalent, that is, 
\be 
\nabla_a V^b= e_a^\m
e^b_\r \nabla_\m V^\r 
\ee 
In this equation we use the spin connection $\omega$ in the left hand side, and the Einstein connection, $\Gamma$ in the right hand side.

It follows that
 \be
\omega_{abc}=-e_c^\r\pd_a e_{b\r}+\eta_{bd}\Gamma^d_{ac}
 \ee
showing that Lorentz and Einstein connections are equivalent
assuming knowledge of the frame field (tetrad).
\par
The Riemann Christoffel tensor is completely analogous to the
usual gauge non-abelian field strength. When the metric compatible connection is used, the main difference
between the curvature tensor and the non-abelian field strength
stems from the torsionless\footnote{Torsion could be easily included; we did not do it mainly for simplicity.}
algebraic Bianchi identity 
\be
R^a\,_b\wedge e^b=0 
\ee 
which is the origin of the symmetry
between Lorentz and Einstein indices 
\be
R_{\a\b\g\d}\equiv e_\a^a
e_\b ^b~R_{a b \g\d}= R_{\g\d\a\b}\equiv e_\g^c e_\d^d R_{c d
	\a\b} 
\ee 

This identity  does not have any analogue in a non
abelian gauge theory in which these two sets of indices remain
unrelated. The opposite happens with the differential Bianchi
identity 
\be 
d R^a\,_b+R^a\,_c\wedge
\omega^c\,_b-\omega^a\,_c\wedge R^c\,_b=0
\ee 
which still holds
for non-abelian gauge theories when the gauge group is not
identified with the tangent group.
\par

The {\em non-metricity tensor} (NM) is just the covariant
derivative of the metric tensor \be \nabla_c \eta_{a b}\equiv
-Q_{
	abc}=-\omega^d\,_{a|c}\eta_{db}-\omega^d\,_{b|c}\eta_{ad}=-\omega_{ba|c}-\omega_{ab|c}
\ee 

Its vanishing characterizes the Levi-Civita connection, whose
components are given by the Christoffel symbols.
The symmetric piece of the connection is then precisely 
\be
\omega_{(ab)|c}= Q_{abc}
\ee
 The structure constants of the frame field are defined by
\be
\left[e_a,e_b\right]\equiv \left[e_a^\m\pd_\m,e_b^\l\pd_\l\right]\equiv C_{ab}^{\quad c}e_c^\s\pd_\s\equiv C_{ab}^c e_c 
\ee

Indeed, the vanishing of the torsion tensor 
\be
de^a+\omega^a\,_b\wedge e^b=0 =\pd_\r e^a_\s -\pd_\s e^a_\r+\omega^a_{\s|\r}-\omega^a_{\r|\s}
\ee
yields the missing antisymmetric
piece of the Lorentz connection $\omega_{a[bc]}$ (remember that
the symmetric piece was determined by the non-metricity)
\be
\omega_{abc}\equiv \omega_{a(cb)}+\omega_{a[bc]}
\ee 
Then
\be
\omega_{[ac]b}+\omega_{[ba]c}+\omega_{[cb]a}={1\over 2}\left(C_{cab}+C_{bca}+C_{abc}\right)
\ee
\be
\omega_{[ab]c}=-{1\over 2} C_{abc}
\ee
The general torsionless
connection is then determined in terms of the non-metricity and
the structure constants of the frame field.
It could be thought
that there is some difference between the use of the one forms 
\be
\omega^a_{ \m b} 
\ee 
(which can be thought of as gauge fields
valued on the Lorentz group O(1,3)) or else the three-index
objects 
\be 
\Gamma^\a_{\m\b} 
\ee
We think this is not the case,
owing to the fact already mentioned, that the Lorentz covariant
derivative is the projection of the Einstein covariant derivative.
\par 
The gauge field $\Gamma\in \mathfrak{gl}(n)$. There is a natural mapping between
\be
\mathfrak{gl}(n)\rightarrow \mathfrak{sl}(n)\times\mathfrak{R}
	\ee
	namely
	\be
	g\rightarrow\left(\hat{g}\equiv  \left(\text{det}~g\right)^{-1/n} g,\quad \text{det}~g\right),
	\ee
	in such a way that any representation of $\mathfrak{gl}(n)$ also yields a representation of $\mathfrak{sl}(n)\times \mathfrak{R}$. The converse is also true.
	Consider a representation $D_k$ of $\mathfrak{R}$
	\be
	r\rightarrow r^k
	\ee
	where $k\in \mathbb{R}$, and a finite-dimensional representation of $\mathfrak{gl}(n)$. This is seen to generate a representation of $\mathfrak{gl}(n)$
	\be
g\quad\rightarrow\quad r^k~\hat{g}
\ee
\par
\subsection{Analogies with a gauge theory}

The fact that the Riemann tensor $R^\m\,_{\n; a b}$ is quite similar to the gauge field strength, $F_{\m\n}$, when viewed as a Lie algebra matrix has been highlighted many times. The thing reads as follows.
Were we to contract in the most natural $SO(n)$ invariant way
\be
R^\m\,_{\n a b}\,g^{a b c d}\,R^\n\,_{\m cd}
\ee
with
\be
g^{abcd}\equiv
\left(\d^{a c}\d^{bd}+\d^{a d}\d^{bc}\right)
\ee
the result is not $GL(n)$ invariant, in spite of the fact that the field A lives in the algebra, $A\in  \mathfrak{gl}(n)$. The reason is, of course, that $g^{abcd}$ is not proportional to the Killing metric of  $\mathfrak{gl}(n)$.The result is only  $\mathfrak{so}(n)$ invariant.
\par
In a gauge theory with gauge group $GL(n)$ the first thing that strikes the eye is that the gauge group  fails to be compact. There is then the general question as to whether any  gauge theory defined with a non-compact version of a given group, is in any sense the analytic continuation of the same theory defined by standard techniques from a compact version of the same group. In \cite{Natan} evidence is given in the negative, at least for three-dimensional Chern-Simons theories.
To be specific, the  $\mathfrak{gl}(4)$ Lie algebra generators are
6 antisymmetric $J_{[\a\b]}$ that generate the $\mathfrak{so}(4)$subgroup, nine traceless symmetric  shears $T_{(\a\b)}$ and one dilatation, $T$. 

\par
The algebra reads \cite{Tomboulis} 
\bea
&\left[J_{\a\b},J_{\g\d}\right]=i\left(\d_{\a\g}J_{\b\d}-\d_{\a\d} J_{\b\g}-\d_{\b\g} J_{\a\d}+\d_{\b\d} J_{\a\g}\right)\nonumber\\
&\left[J_{\a\b},T_{\g\d}\right]=i\left(\d_{\a\g}T_{\b\d}+\d_{\a\d} T_{\b\g}-\d_{\b\g} T_{\a\d}-\d_{\b\d} T_{\a\g}\right)\nonumber\\
&\left[T_{\a\b},T_{\g\d}\right]=i\left(\d_{\a\g}T_{\b\d}+\d_{\a\d} T_{\b\g}+\d_{\b\g} T_{\a\d}+\d_{\b\d} T_{\a\g}\right)
\eea
The  $\mathfrak{so}(3)\oplus \mathfrak{so}(3)$ subalgebra can be highlighted by defining
\bea
&J_1\equiv J_{23}\quad\quad  K_1\equiv J_{14}\nonumber\\
&J_2\equiv -J_{13} \quad\quad K_2\equiv J_{24}\nonumber\\
&J_3\equiv J_{12}\quad\quad K_3\equiv J_{34}
\eea
as well as
\bea
&J_i^\pm\equiv{J_i \pm K_i\over 2}
\eea
Then the first line of the algebra collapses to
\bea
&\left[J^\pm_i,J^\pm_j\right]=i\,\e_{ijk}\, J_\pm^k\nonumber\\
&\left[J^\pm_i,J^\mp_j\right]=0
\eea
The symmetric generators (which do not close in a subgroup) can be thought of as  ${n(n-1)\over 2}$ non diagonal traceless matrices, plus $n-1$ diagonal traceless ones; plus the trace. An explicit representation is
\bea
&\left(J_{\a\b}\right)_{\r\s}\equiv \d_{\a\r}\d_{\b\s}-\d_{\a\s}\d_{\b\r}
\nonumber\\
&\left(T_{\a\b}\right)_{\r\s}\equiv \d_{\a\r}\d_{\b\s}+\d_{\a\s}\d_{\b\r}
\nonumber\\
&\left(T^D_{\a\b}\right)_{\r\s}\equiv \d_{\a\r}\d_{\a\s}+\d_{\b\s}\d_{\b\r}\nonumber\\
&T_{\r\s}\equiv \d_{n\r}\d_{n\s}
\eea
It is a known fact that the Killing form of $\mathfrak{gl}(n)$ is given by
\be
B(A,B)\equiv 2\left( n\, \text{tr}\, (AB)-\text{tr} A\, \text{tr}\, B\right)
\ee
When we put indices
\be
B^{abcd}\equiv 2\left( n\d^{bc}\d^{ad}-\d^{ab}\d^{cd}\right)
\ee
The generator responsible for the group not being semisimple is just the {\em dilatation} 
\be
T\sim t\,  I
\ee
 because 
\be
B(T,T)=0
\ee
 whereas the remaining traceless generators of $\textfrak{gl(n)}$ are responsible for non-compactedness even when the algebra belongs to the first $A_n$-Cartan series.
\par
It is well-known \cite{Natan}\cite{Tseytlin} that when the gauge group  is non compact (which manifests itself in the Killing metric not being positive definite, actually of signature $\left({n(n-1)\over 2},{(n-1)(n+2)\over 2}\right)$, some analytic continuation is in order (which naively means putting all euclidean signs as $+$). As pointed out in those references sometimes (like in the Chern-Simons case) even that is not enough and some more elaborate physical analysis is in order.
\par
The relationship between $A_\m$ and $\omega_\m$
is
\be
\omega_\m^a\,_b=\Gamma^a_{\m b}-e_b^\l \pd_\m e^a_\l
\ee
The Einstein index in $\Gamma_\m$ is not the contravariant  one, but rather one of the two equivalent  covariant ones. It is quite easy to check that
\bea
&\omega_{\m[ab]}\equiv \Gamma_{[a|\m|b]}-e_{[b}^\l\pd_\m e_{a]|\l}\nonumber\\
&\omega_{\m(ab)}=\Gamma_{(a|\m|b)}\equiv{1\over 2}\left(\Gamma_{a\m b}+\Gamma_{b\m a}\right)
\eea
 where the  antisymmetry instruction acts on the $a,b$ indices only.
 All these results allow to trade the $\mathfrak{so(n)}$ metric for the $\mathfrak{gl(n)}$ metric if so desired. As it stands, the theory is only invariant under an $\mathfrak{so(n)} \subset \mathfrak{gl(n)}$ subgroup, and the full gauge symmetry is broken by the kinetic energy metric.
 \par
The absence of torsion implies
\be
\Gamma_{a|\m b}=\Gamma_{a|b\m}
\ee
Actually, the {\em difference} between an arbitrary connection and
the Levi-Civita one
 \bea
A^\a_{\m \b}\equiv  \Gamma^\a_{\m \b} -\footnotesize{\left\{\begin{array}{c}
\a \\ 
\m ~ \b
\end{array}\right\}}
\eea
is a true tensor, so that
\be
 A^\a_{\m\b}\equiv e^\a_a~ A^a_{\m b}~ e^b_\b
  \ee
   and there is
a simple field redefinition between the two languages. The condition we have imposed of absence of torsion has a much simpler expression in spacetime language (where it just states that the connection is symmetric) than in the frame one, where it reads
\be
\omega_{u|av}-\omega_{v|a u}=e^\s_v \pd_u e_{a\s}-e^\r_u\pd_v e_{a\r}
\ee
\par

As will be seen soon, the most general lagrangian is a quite complicated one, with  12
independent coupling constants and many possible vacua to
consider. In this paper we present the general setup and analyze
the response of the  flat vacuum to external graviton sources.

\par
To be specific, in the second section we analyze the case of General relativity in the first order formalism. In  section three we do a careful study of the independent monomia that can be written with the assumed fields and symmetries, and we find that there are indeed twelve of them. In the fourth section a background field expansion is performed. Many unwieldy formulas are relegated to an appendix. Then we study the effect of external sources on the system, and we analyze  carefully the conditions for this effect to mimic the one of General Relativity in section five. The necessity to break Weyl invariance in order to make contact with phenomenology is emphasized in section six.
Finally, we end this work with some conclusions.
\par
A word of warning. We shall still call the metric fluctuations, $h_{\m\n}$ {\em graviton fluctuations} and the fluctuations of the connection, $A_{\a\b\g}$ {\em (three-index) gauge fluctuations}, in spite of the fact that both are related to the gravitational field.

\section{General relativity.}
In order to understand the role of external sources in first order formalism, let us consider first the Einstein-Hilbert action (FOEH).
\par
\noindent To be specific, we define the action like
\be
 S_\text{FOEH}\equiv-{1\over 2 \kappa^2}~\int
d^n x\sqrt{|g|}g^{\m\n}R_{\m\n}
\ee
On the one hand, it is well known that the classical equations of motion are equivalent to Einstein's equations. Our aim here is to understand this from the path integral in the presence of external sources. The first question is, which sources? In principle, we are supposed to assume sources for physical fields only. This would mean to include a source for the graviton field , and not for the connection. We shall come back to that.
\par
\subsection{A toy model.}
In order to understand properly what is going on, let us first consider an ordinary integral that shares most of the features of our path integral, namely,
\be
I(j,h)\equiv \int_{-\infty}^\infty dx dy~e^{-n x y - k y^2 -j x -h y}
\ee
Let us first compute $I(0,0)$ in two different ways. We shall as usual, define the integrals by analytic continuation from the region where they are convergent. First, complete the square
\be
-n x y - k y^2= -k\left(y + {n x \over 2 k}\right)^2- {n^2 x^2\over 4 k}
\ee
It follows that
\be
I(0,0)=\sqrt{\pi\over k}\int_{-\infty}^\infty dx~e^{-{ n^2 x^2 \over 4 k}}=\sqrt{\pi\over k}\sqrt{ 4 \pi \over n^2/k}={2\pi\over n}
\ee
A different way to proceed would be to first perform the integral over $dx$, getting 
\be
2 \pi \d( n y)={2\pi\over n}\d(y)
\ee
The integral over $dy$ is now immediate, yielding again
\be
I(0,0)={2\pi\over n}
\ee
In the presence of sources, the integral over $dx$ yields
\be
2\pi \d(n y + j)={2\pi \over n}~\d(y+{j\over n})
\ee
so that
\be
I(j,h)={2\pi\over n}~e^{-{jkj\over n^2}+{hj\over n}}
\ee

\subsection{Einstein-Hilbert in first order.}
Let us start by analyzing the action $S_{\text{FOEH}}$ with a graviton source 
\be S_M=-\frac{1}{2}\int
d^4 x~\kappa h^{\g\e}T_{\g\e}\ee
we expand around Minkowski spacetime as
\bea
&g_{\m\n}\equiv \eta_{\m\n}+\kappa h_{\m\n}\nonumber\\
&\Gamma^\a_{\b\g}\equiv  A^\a_{\b\g}
\eea
this yields

\bea S_{FOEH}+S_M&=\bar{S}_0-\int
d^n x\left\{\frac{1}{2}\left(h^{\g\e}N_{\g\e~\l}^{~~\a\b}A^{\lambda}_{\alpha\beta}+A^{\lambda}_{\alpha\beta}N_{~\l~\g\e}^{\a\b}h^{\g\e}\right)+\right.\nonumber\\
&\left.+\frac{1}{2}A^{\t}_{\g\e}K^{\g\e~\a\b}_{~\t~~\l}A^{\lambda}_{\alpha\beta}+\frac{1}{2}\kappa h^{\g\e}T_{\g\e}\right\}\nonumber\\
\eea
where 
\be \bar{S}_0=-{1\over 2 \kappa^2}~\int
d^n x\sqrt{|\bg|}\bR\ee
and
\bea
N_{\g\e~\l}^{~~\a\b}&=\frac{1}{2\kappa}\left\{\frac{1}{2}\left(\eta_{\g\e}\eta^{\a\b}-\d^\a_\g\delta^{\beta}_{\e}-\d^\a_\e\delta^{\beta}_{\g}\right)\bar{\nabla}_\lambda
-\right.\nonumber\\
&\left.-\frac{1}{4}\left(\eta_{\g\e}\d^\b_\l\bar{\nabla}^\a -\d^\a_\g\delta^{\beta}_{\lambda}\bar{\nabla}_\e-\d^\a_\e\delta^{\beta}_{\lambda}\bar{\nabla}_\g
+\eta_{\g\e}\d^\a_\l\bar{\nabla}^\b -\d^\b_\g\delta^{\a}_{\lambda}\bar{\nabla}_\e-\d^\b_\e\delta^{\a}_{\lambda}\bar{\nabla}_\g\right)\right\}\nonumber\\
K^{\g\e~\a\b}_{~\t~~\l}&=\frac{1}{\kappa^2}\left\{\frac{1}{4}\big[\d^\e_\t \d^\g_\l \eta^{\a\b}
+\d^\g_\t \d^\e_\l \eta^{\a\b}
-\d^\b_\t \d^\g_\l \eta^{\a\e}
-\d^\b_\t \d^\e_\l \eta^{\a\g}
-\d^\a_\t \d^\e_\l \eta^{\b\g}
-\d^\a_\t \d^\g_\l \eta^{\b\e}+\right.\nonumber\\
&\left.+\d^\b_\l \d^\a_\t \eta^{\g\e}
+\d^\a_\l \d^\b_\t \eta^{\g\e}
\big]\right\} \label{K}
\eea
Let us define as usual 
\be\label{Carmel}\displaystyle
Z[T]= \displaystyle\int {\cal D} \varphi~e^{\left\{i S[\varphi]+i\displaystyle\int d^n x~T(x)\varphi(x)\right\}}
\ee  
so that the  free energy, 
\be e^{iW}=\frac{Z[T]}{Z[0]}\ee
reads in our case
\bea
e^{iW_{\text{FO}}\left[T_{\m\n}\right]}=\displaystyle\int \mathcal{D}h\mathcal{D}A e^{\left\{-i\int
	d^n x~\Big(\frac{1}{2}\left(h^{\g\e}N_{\g\e~\l}^{~~\a\b}A^{\lambda}_{\alpha\beta}+A^{\lambda}_{\alpha\beta}N_{~\l~\g\e}^{\a\b}h^{\g\e}\right)+\frac{1}{2}A^{\t}_{\g\e}K^{\g\e~\a\b}_{~\t~~\l}A^{\lambda}_{\alpha\beta}+\frac{1}{2}\kappa h^{\g\e}T_{\g\e}\Big)\right\}}\nonumber\\\eea

Our purpose in life is  to derive the lowest order interaction between external sources.
Notice that the quadratic graviton term
\be
\int h \overline{M} h
\ee
vanishes in our case.

Let us face the consequences of this fact.
Integrating over ${\cal D} h$ yields a Dirac delta
\be
\d(\widebar{N}_{\g\e~\l}^{~~\a\b}A^{\lambda}_{\alpha\beta}+\kappa T_{\g\e})
\ee
we define by $\widebar{A}$ the solution of the equation
\be
\widebar{N}_{\g\e~\l}^{~~\a\b}\widebar{A}^{\lambda}_{\alpha\beta}=-\kappa T_{\g\e}\label{T}
\ee
This is not an EM for any background field; it is the argument of a Dirac delta function, consequence of having integrated ${\cal D}h_{\m\n}$ away.
\par
Then it is clear that (modulo a jacobian independent of the sources) the integral over ${\cal D}A$ yields
\be
W_{FOEH}[T_{\m\n}]=-\frac{1}{2} \int
d^n x~\widebar{A}^{\t}_{\g\e}\widebar{K}^{\g\e~\a\b}_{~\t~~\l}\widebar{A}^{\lambda}_{\alpha\beta}+\log~J
\ee
and this should be proportional to $W_{SOEH}[T_{\m\n}]$ \eqref{jes}.

\bea W_{SOEH}\left[T_{\m\n}\right]&=-\int
d^n x~\frac{\kappa^2}{4k^2}T_{\m\n}\left(\eta^{\m\r}\eta^{\n\s}+\eta^{\m\s}\eta^{\n\r}-\eta^{\m\n}\eta^{\r\s}\right)T_{\r\s}\eea
then
\be \widebar{A}^{\t}_{\g\e}\widebar{K}^{\g\e~\a\b}_{~\t~~\l}\widebar{A}^{\lambda}_{\alpha\beta}=\frac{\kappa^2}{2k^2}T_{\m\n}\left(\eta^{\m\r}\eta^{\n\s}+\eta^{\m\s}\eta^{\n\r}-\eta^{\m\n}\eta^{\r\s}\right)T_{\r\s}\label{1}
\ee
Let us now determine $\widebar{A}$. In momentum space (\ref{T})
\bea
\frac{1}{\kappa}\left\{\frac{1}{2}\eta_{\g\e}\left(\eta^{\a\b}k_\lambda \bar{A}^{\lambda}_{\alpha\beta}-k^\a \bar{A}^{\lambda}_{\alpha\l}\right)
+\frac{1}{2}\left(k_\e \bar{A}^{\lambda}_{\g\l}+k_\g \bar{A}^{\lambda}_{\e\l}-2k_\lambda \bar{A}^{\lambda}_{\g\e}\right)\right\}=-\kappa T_{\g\e}
\eea
The integrability condition stemming from conservation of the source
\be
k_\m T^{\m\n}=0
\ee
necessary for maintaining gauge invariance determines uniquely
\be \bar{A}^\l_{\a\b}= f(k^2)\left[k^\l \eta_{\a\b}-\left(\d^\l_\a k_\b+\d^\l_\b k_\a\right)\right]\ee
Assuming, that is, that it depends on the metric and the momentum only. In this same spirit, the source must be
\be T_{\m\n}=-\frac{2-n}{\kappa^2} f(k^2)[k_\m k_\n-k^2\eta_{\m\n}]\ee

With these expressions of $\bar{A}^\l_{\a\b}$ and $T_{\m\n}$, we can work out the equation  (\ref{1})
\be 2(2-n)(n-1)=(3-n)(2-n)^2(n-1)\ee
This equation admits $n=4$ as a solution.

We can integrate instead over  the connection perturbation, yielding
\bea
e^{iW_{\text{FO}}\left[T_{\m\n}\right]}=\int \mathcal{D}h e^{\left\{-i\int
	d^n x(\frac{1}{2}h^{\m\n}N_{\m\n~\l}^{~~\a\b}(K^{-1})_{\a\b~\g\e}^{~\l~~\t}N_{~\t~\r\s}^{\g\e}h^{\r\s}+\frac{1}{2}\kappa h^{\g\e}T_{\g\e})\right\}}\eea
where the graviton operator is
\bea
N_{\m\n~\l}^{~~\a\b}(K^{-1})_{\a\b~\g\e}^{~\l~~\t}N_{~\t~\r\s}^{\g\e}&=\frac{1}{8}(\eta_{\m\r}\eta_{\n\s}+\eta_{\m\s}\eta_{\n\r}-2\eta_{\m\n}\eta_{\r\s})\Box-\nonumber\\
&-\frac{1}{8}(\eta_{\m\r}\partial_\n \partial_\s+\eta_{\m\s}\partial_\n \partial_\r-2\eta_{\m\n}\partial_\r \partial_\s+\eta_{\n\r}\partial_\m \partial_\s+\eta_{\n\s}\partial_\m \partial_\r-2\eta_{\r\s}\partial_\m \partial_\n)\nonumber\\
\eea
It is easy to check that this whole action is invariant under the gauge symmetry
\bea
&\d h_{\m\n}=\pd_\m\xi_\n+\pd_\n\xi_\m\nonumber\\
&\d A^\a_{\b\g}=\pd_\b\pd_\g\xi^\a
\eea
that we need to fix.
\par
Therefore, we still have the freedom to fix the gauge in a way that simplifies the computation.
The gauge fixing term will be
\be
S_{\text{gf}}=\frac{1}{2}\,\,\int\,d^nx\,\,\,\frac{1}{2\xi}\,\eta_{\m\n}\chi^\m\chi^\n
\ee
where the function characterizing the harmonic gauge is
\be
\chi^\n=\partial_\m h^{\m\n}-\frac{1}{2}\partial^\n h
\ee	
and in the minimal gauge, corresponding to $\xi=1$, the path integral can be rewritten as
\bea
e^{iW^{\text{gf}}_{\text{FO}}\left[T_{\m\n}\right]}=\int \mathcal{D}h e^{\left\{-\frac{i}{2}\int
	d^n x~~(-\frac{1}{2}h^{\m\n}D_{\m\n\r\s} h^{\r\s}+\kappa h^{\g\e}T_{\g\e})\right\}}\eea
where
\bea
D_{\m\n\r\s}&=\frac{1}{4}\left(\eta_{\m\l}\eta_{\n\s}+\eta_{\m\s}\eta_{\n\l}- \eta_{\m\n}  \eta_{\l\s}\right)\Box\label{DFOEH}
\eea
Finally, we can integrate over $h$
\bea
e^{iW^{\text{gf}}_{\text{FO}}\left[T_{\m\n}\right]}= e^{\left\{-\frac{i}{4}\kappa^2\int
	d^n x~T^{\m\n}D_{\m\n\r\s}^{-1}T^{\r\s}\right\}}\eea
Getting the result,
\bea
W^{gf}_{FO}\left[T_{\m\n}\right]=-\frac{\kappa^2}{4}\int
d^n x~~T^{\m\n}D_{\m\n\r\s}^{-1}T^{\r\s}\eea
It is remarkable that the divergent part also coincides exactly {\em off-shell} \cite{Anero}.
\subsection{Einstein-Hilbert in second order.}
Now, we consider the Einstein-Hilbert action  in second order.  In the same way that before, we perform an expansion around flat space $g_{\m\n}=\eta_{\m\n}+\kappa h_{\m\n}$. This reads
\bea S_{SOEH}+S_M&=-\frac{1}{2}\int
d^4 x~\left\{\frac{1}{2}\left(\frac{1}{2}\partial_\l h^{\m\n}\partial^\l h_{\m\n}-\frac{1}{2}\partial_\l h \partial^\l h-\partial_\l h^{\l\n}\partial^\m h_{\m\n}+\partial^\n h\partial^\m h_{\m\n}\right)+\kappa h^{\g\e}T_{\g\e}\right\}\nonumber\\\label{SOEH}
\eea
Adding the usual harmonic gauge fixing $\partial_\m h^\m_{\n}=\frac{1}{2}\partial_\n h^\l_{\l}$, and integrating by parts in (\ref{SOEH}) we get
\bea S^{gf}_{SOEH}+S_M&=-\frac{1}{2}\int
d^4 x~\left\{-\frac{1}{4}\left(h^{\m\n}\partial^2 h_{\m\n}-\frac{1}{2} h \partial^2 h\right)+\kappa h^{\g\e}T_{\g\e}\right\}=\nonumber\\
&=-\frac{1}{2}\int
d^4 x~\left\{-\frac{1}{2}h^{\m\n}\frac{1}{4}\left(\eta_{\m\l}\eta_{\n\s}+\eta_{\m\s}\eta_{\n\l}- \eta_{\m\n}  \eta_{\l\s}\right)\partial^2 h^{\l\s}+\kappa h^{\g\e}T_{\g\e}\right\}
\eea 
that we can again rewrite as
\bea S^{gf}_{SOEH}+S_M&=-\frac{1}{2}\int
d^4 x~\left\{-\frac{1}{2} h^{\a\b}D_{\a\b\g\e}h^{\g\e}+\kappa h^{\g\e}T_{\g\e}\right\}
\eea
This is the same operator that in \eqref{DFOEH}, so we get for the free energy
\bea\label{jes}
 W^{\text{gf}}_{\text{SO}}&=-\frac{\kappa^2}{4}\int
d^n x~\sqrt{|g|}T^{\m\n}D^{-1}_{\m\n\r\s}T^{\r\s}\eea
The final conclusion is that the first order formalism is equivalent to the second order one with external sources for the graviton
$W^{\text{gf}}_{\text{FO}}=W^{\text{gf}}_{\text{SO}}$.

This seems the best procedure in order to compute the one loop divergences by heat kernel methods.

\newpage
\section{ The most general quadratic action.}
\subsection{First order versus second order}

In the paper \cite{Borunda} a full analysis is made of first order versus second order EM and it is concluded that coincidence in the above sense (that is, once the Levi-Civita connection has been substituted in the general EM) is only found for Lanczos-Lovelock (LL) and related lagrangians.
\par
Anticipating the notation we shall introduce in our equation \eqref{lo} this happens when
\bea
&\a_1=\a_3=-{\a_2\over 4}
\eea
It is of course well-known that quadratic LL lagrangians are trivial in four dimensions (where they reduce to the Gauss-Bonnet density), but they appear in brane-world scenarios as well as in some dark matter proposals.
There are other, less restrictive, instances where the EM are also equivalent in the above sense. The starting point is  the equation found in  \cite{Borunda} giving the difference between both EM, namely
\bea\label{Bert}
&\Delta\,H_{\m\n}\equiv  H_{\m\n}^{SO}-H_{\m\n}^{FO}=-{1\over 2}\nabla_{\l} K^\l_{(\m\n)}+{1\over 4} g_{\l\m}\,\nabla^\r K^\l_{\r\n}+{1\over 4} g_{\l\n}\, \nabla^\r K_{\r\n}
\eea
where
\be
H_{\m\n}\equiv {1\over \sqrt{|g|}}\,{\d S\over \d g^{\m\n}}
\ee
and
\bea
&{1\over \sqrt{|g|}}~{\d S \over \d \Gamma^{\m\n}_\l}\,\equiv K^\l_{\m\n}\equiv 2(\a_2+\a_3) g_{\m\n}\nabla^\l R+(\a_2+4\a_1)\nabla^\l R_{\m\n}-\nonumber\\
&-2(\a_3+\a_2)\d^\l_\n\,\nabla_\m R-(\a_1+\a_2)\nabla_\m R_\n^\l
\eea
It is plain that for constant curvature backgrounds the whole tensor $K^\l_{\m\n}$ vanishes and both sets of EM are equivalent.
\par
Actually more is true. In this same reference \cite{Borunda} general lagrangians involving the metric and the Riemann tensor (but not its derivatives) have been considered.
\be
S=\int \sqrt{|g|} d^n x\,L\left(g_{\a\b},R^\m\,_{\n\r\s}\right)
\ee
Again, in the Levi-Civita case, the relationship between the first order and second order  EM is {\em exactly} as in \eqref{Bert}, and besides,
\be
K_\l^{\m\r}=\nabla_\n B_\l^{\n\m\r}
\ee
with
\be
B_\l^{\m\n\r}={\d L\over \d R^\l\,_{\m\r\n}}-{\d L\over \d R^\l\,_{\m\n\r}}
\ee
\subsection{Quadratic actions}
It is worth pointing  out that  when the nonmetricity is
non-vanishing the Riemann tensor does not enjoy the usual
symmetries
\bea &R[\Gamma]_{\m\n\r\s}\neq R[\Gamma]_{\r\s\m\n}\\  &R[\Gamma]_{(\m\n)\r\s}\neq 0 
\eea There are then two
different traces. The one that corresponds to the Ricci tensor 
\be
R^+[\Gamma]_{\n\s}\equiv g^{\m\r}~R[\Gamma]_{\m\n\r\s} \label{ricciplus}\ee 
and a different one 
\be R^-[\Gamma]_{\m\s}\equiv
g^{\n\r}R[\Gamma]_{\m\n\r\s}\label{ricciminus} \ee 
Neither of them is in general
symmetric now. There is also an antisymmetric further trace
\be\label{ricciblando} {\cal R}[\Gamma]_{\r\s}\equiv
g^{\m\n}~R[\Gamma]_{\m\n\r\s} \ee 
but it is easy to check that
\be {\cal R}[\Gamma]_{\r\s}\equiv
R^+[\Gamma]_{\r\s}-R^+[\Gamma]_{\s\r} \ee 
However, there is an only scalar \be R^+\equiv
g^{\m\n} R^+_{\m\n}=-R^-\equiv g^{\m\n} R^-_{\m\n}\label{traces}
\ee

while $g^{\r\s}{\cal R}_{\r\s}=0$.\par

Let us now  write the most general quadratic action, made with 12 Weyl scalars.\footnote{ We only consider parity-conserving operators, therefore terms like 
	\be \e^{\m\n\r\s}R^{\pm}_{\m\n}R^{\pm}_{\r\s},~~\e^{\m\n\r\s}R^{\pm}_{\m\n}R^{\mp}_{\r\s},~~\e^{\m\n\r\s}\cR_{\m\n}\cR_{\r\s}\ee
are excluded.}
\par
There are then five independent quadratic scalar operators that can be built out of two Riemann tensors
which are all of the general form
\be\label{cano}
\cO_I \equiv ~R^\m\,_{\n\r\s}
~\left(D_I\right)_{\m\m^\prime}^{\n\r\s\n^\prime\r^\prime\s^\prime}~R^{\m^\prime}\,_{\n^\prime
	\r^\prime\s^\prime}
\ee
for $I=1\ldots 5$, where
\begin{table}[H]
	\begin{center}
		\begin{tabular}{l | l }
			$(D_1)_{\m\m'}^{\n\n'\rho\rho'\s\s'}=g_{\m\m'}g^{\n\n'}g^{\r\r'}g^{\s\s'}\nonumber$&\qquad$\cO_1~ \equiv~\sqrt{|g|}~R_{\m\n\r\s}~R^{\m\n\r\s}$\\
			$(D_2)_{\m\m'}^{\n\n'\rho\rho'\s\s'}=g_{\m\m'}g^{\n\r'}g^{\r\n'}g^{\s\s'}$ &\qquad $\cO_2~ \equiv~\sqrt{|g|}~R_{\m\n\r\s}~R^{\m\r\n\s}$\\
			$(D_3)_{\m\m'}^{\n\n'\rho\rho'\s\s'}=\d^{\n'}_{\m}\d^{\n}_{\m'}g^{\r\r'}g^{\s\s'}$&\qquad$\cO_3~ \equiv~\sqrt{|g|}~R_{\m\n\r\s}~R^{\n\m\r\s}$\\
			$(D_4)_{\m\m'}^{\n\n'\rho\rho'\s\s'}=\d^{\r'}_{\m}\d^{\r}_{\m'}g^{\n\n'}g^{\s\s'}$&\qquad$\cO_4~ \equiv~\sqrt{|g|}~R_{\m\n\r\s}~R^{\r\n\m\s}$\\
			$(D_5)_{\m\m'}^{\n\n'\rho\rho'\s\s'}=\d^{\r'}_{\m}\d^{\r}_{\m'}g^{\n\s'}g^{\s\n'}$&\qquad$\cO_5~ \equiv~\sqrt{|g|}~R_{\m\n\r\s}~R^{\r\s\m\n}$
		\end{tabular}
	\end{center}
\end{table}
We follow the Landau-Lifshitz spacelike conventions, in particular
\be
R^\m\,_{\n\r\s}= \pd_\r\Gamma^\m_{\n\s}-\pd_\s\Gamma^\m_{\n\r}+ \Gamma^\m_{\l\r}\Gamma^\l_{\n\s}-\Gamma^\m_{\l\s}\Gamma^\l_{\n\r} 
\ee
A remarkable fact is that under Weyl rescaling
\be
g_{\m\n}\rightarrow \Omega^2(x) g_{\m\n}
\ee
assuming the connection remains inert, all operators transform as
\be
\cO_I\rightarrow \Omega^{n-4}~\cO_I
\ee
so that all these operators remain Weyl invariant when integrated in four dimensions. This means that the most general quadratic action is Weyl invariant in this sense.
\par
There are then six Weyl scalar operators that can be formed with the two different traces, (\ref{ricciplus}) and (\ref{ricciminus}) 
\begin{table}[H]
	\begin{center}
		\begin{tabular}{ll | l }
			&$(D_6)_{\m\m'}^{\n\n'\rho\rho'\s\s'}=\d^{\r}_{\m}\d^{\r'}_{\m'}g^{\n\n'}g^{\s\s'}$ &\qquad $\cO_6\equiv~ \sqrt{|g|}~ R^+_{\m\n}~R_+^{\m\n}$\\
			& $(D_7)_{\m\m'}^{\n\n'\rho\rho'\s\s'}=\d^{\r}_{\m}\d^{\r'}_{\m'}g^{\n\s'}g^{\s\n'} $&\qquad $\cO_7\equiv ~ \sqrt{|g|}~R^+_{\m\n}~R_+^{\n\m}$\\
			&$(D_8)_{\m\m'}^{\n\n'\rho\rho'\s\s'}=g_{\m\m'}g^{\n\r}g^{\n'\r'}g^{\s\s'}$ &\qquad $\cO_{8}\equiv~ \sqrt{|g|}~ R^-_{\m\n}~R_-^{\m\n}\nonumber $\\
			& $(D_{9})_{\m\m'}^{\n\n'\rho\rho'\s\s'}=\d^{\s'}_{\m}\d^{\s}_{\m'}g^{\n\r}g^{\n'\r'}$ &\qquad $\cO_{9}\equiv ~ \sqrt{|g|}~R^-_{\m\n}~R_-^{\n\m}\nonumber $\\
			&$(D_{10})_{\m\m'}^{\n\n'\rho\rho'\s\s'}=\d^{\r}_{\m}\d^{\n}_{\m'}g^{\n'\r'}g^{\s\s'} $&\qquad $\cO_{10}\equiv ~ \sqrt{|g|}~R^+_{\m\n}~R_-^{\m\n} $\\
			&$(D_{11})_{\m\m'}^{\n\n'\rho\rho'\s\s'}=\d^{\r}_{\m}\d^{\s}_{\m'}g^{\n'\r'}g^{\n\s'} $&\qquad $\cO_{11}\equiv~ \sqrt{|g|}~ R^+_{\m\n}~R_-^{\n\m}$\\
		\end{tabular}
	\end{center}
\end{table}

\par
Finally there is only one independent curvature scalar operator
\be
\cO_{12}\equiv \sqrt{|g|}~R^2
\ee
which also admits the canonical form (\ref{cano}) with
\be
\left(D_{12}\right)_{\m\m^\prime}^{\n\r\s\n^\prime\r^\prime\s^\prime}\equiv \d^\r_\m ~\d^{\r^\prime}_{\m^\prime}~g^{\n\s}~g^{\n^\prime \s^\prime}
\ee
The most general Weyl invariant lagrangian is then a sum of these twelve operators with arbitrary coefficients
\be
\cL\equiv \sum_{I=1}^{I=12}~g_I~\cO_I
\ee
where $g_I$ are arbitrary, generically non-vanishing, dimensionless coupling constants. This lagrangian is expected to be renormalizable by power counting. Unitarity may be an issue and has to be analyzed in detail.
\par
Finally, we can write the most general quadratic action as
\be S=\int
d^n
x\sqrt{|g|}\cL\left(g_{\mu\nu},A^{\tau}_{\gamma\epsilon}\right)=\int d^n x\sqrt{|g|}\sum_{I=1}^{I=12}g_I
R^{\mu}_{\quad\nu\rho\sigma}(D_I)_{\mu\mu'}^{\nu\nu'\rho\rho'\sigma\sigma'}R^{\mu'}_{\quad\nu'\rho'\sigma'}\label{Fl}\ee

The mass dimension of all coupling constants is
\be
\left[g_I\right]=n-4
\ee
\newpage
\section{ Background field expansion.}
The general background field expansion reads \bea
&g_{\m\n}=\bg_{\m\n}+\kappa h_{\m\n}\nonumber\\
&\Gamma^\m_{\n\r}=\bG^\m_{\n\r}+ B^\m_{\n\r}\equiv
\footnotesize{\left\{\begin{array}{c}
	\m \\ 
	\n~\r
\end{array}\right\}}+A^\m_{\n\r} \eea 
that is, we can
assume without loss of generality that the background connection
$\bG^\m_{\n\r}$ is the Levi-Civita connection corresponding to the
metric $\bg_{\m\n}$.  The tensor $A^\m_{\n\r}$ contains all the
relevant information on the non-metricity of the connection.\footnote{We use a torsionless connection, i.e. $A^\m_{\n\r}=A^\m_{\r\n}$ (for an analysis of quadratic theories with torsion see \cite{Hehl,Shapiro}).

Observe that this is a perfectly acceptable expansion. Were we to allow torsion,  $A^\l_{\a\b}-A^\l_{\b\a}=T^\l_{\a\b}$, the connection tadpole will read
\bea \left.\delta S\right|_{g_{\mu\nu}=\bar{g}_{\mu\nu}}&=\int
d^n x~\sqrt{|\bg|}\sum_{I=1}^{I=12}g_I
R^{\mu}_{\quad\nu\rho\sigma}(D_I)_{\mu\mu'}^{\nu\nu'\rho\rho'\sigma\sigma'}\times\nonumber\\
&\times\left\{
\d^{\m'}_{\l'}\left(\d^{(\b'}_{\n'}\d^{\a')}_{\s'}\bar{\nabla}_{\rho'}-\d^{(\b'}_{\n'}\d^{\a')}_{\r'}\bar{\nabla}_{\s'}\right)+
\d^{\m'}_{\l'}\left(\d^{[\b'}_{\n'}\d^{\a']}_{\s'}\bar{\nabla}_{\rho'}-\d^{[\b'}_{\n'}\d^{\a']}_{\r'}\bar{\nabla}_{\s'}\right)\right\}A^{\lambda'}_{\alpha'\beta'}
\eea
Therefore the torsionless choice is allowed as it corresponds to the case where the second term is zero.} 
The constant $\kappa $ has mass dimension
\be
\left[\kappa\right]=1-\frac{n}{2}
\ee
adequate for the kinetic energy of the field $h_{\m\n}$ to be canonically normalized (that is, $\left[h_{\m\n}\right]={n\over 2}-1$).
This means that in spite of the fact that 
\be \nabla_\r
g_{\m\n}\neq 0 
\ee 
which prevents integration by parts 
\be \int
\sqrt{|g|} d^n x \nabla_\m V^\m \neq 0
\ee 
we can always write \be
\int ~\sqrt{\bg}~ d^n x~ \nabla_\m V^\m =\int ~\sqrt{\bg}~ d^n x
~\left(\bn_\m V^\m+A^\l_{\l\m}~ V^\m\right)=\int ~\sqrt{\bg}~d^n
x~A^\l_{\l\m} V^\m \ee 
Therefore integration  by parts is
still possible at the price of introducing potential terms
involving the field $A^\l_{\m\n}$. We shall then continue using
the notation \be d(vol)\equiv \sqrt{\bg}~d^n x \ee when
appropriate. Let us define \bea
&A^\l\equiv g^{\m\n} A^\l_{\m\n}\nonumber\\
&A_\s\equiv A^\l_{\s\l} \eea Please note that \be A^\l\neq
g^{\l\m} A_\m \ee 
It is also natural to define a {\em field strength} and a quadratic term 
\bea &F^\m_{~\n\r\s}\equiv \bn_\r A^\m_{\n\s}-\bn_\s
A^\m_{\n\r} \\
&O^\m_{~\n\r\s}\equiv A^\m_{\l\r}A^\l_{\n\s}-A^\m_{\l\s}A^\l_{\n\r}
\eea but there is  an extra symmetry, similar to the usual algebraic Bianchi identity
\bea &F^\m_{~\n\r\s}{+} F^\m_{~\s\n\r}+F^\m_{~\r\s\n}=0\\
&O^\m_{~\n\r\s}{+} O^\m_{~\s\n\r}+O^\m_{~\r\s\n}=0 \eea

In this way the  Riemann tensor reads \be
R^\m\,_{\n\r\s}=\widebar{R}^\m\,_{\n\r\s}+F^\m_{~\n\r\s}+
O^\m_{~\n\r\s} \ee where the first term is just the contribution
of the background; the second is linear in the connection
fluctuations, and the third is quadratic in the same quantities.
\par
We can define two different traces for $F^\m_{~\n\r\s}$ and $O^\m_{~\n\r\s}$, in a manner identical to the
way we did it for Riemann's tensor \bea
&F^+_{\n\s}\equiv F^\l_{~\n\l\s}=\nabla_\l A^\l_{\n\s}-\nabla_\s A^\l_{\n\l}\nonumber\\
&F^{\m}_{-\s}\equiv g^{\n\r} F^\m_{~\n\r\s}=\nabla^\l A^\m_{\l\s}-g^{\n\r}\nabla_\s A^\m_{\n\r}\nonumber\\
&O^{+}_{\n\s}\equiv O^\l_{\n\l\s}=A_{\a}A^\a_{\n\s}-A^\l_{\a\s}A^\a_{\n\l}\nonumber\\
&O^{\m}_{-\s}\equiv g^{\n\r}O^\m_{\n\r\s}=g^{\n\r}A^\m_{\l\r}A^\l_{\n\s}-A^\m_{\l\s}A^\l\nonumber\\
 \eea It should be noted that
this objects are not symmetric in general \be F^\pm_{\n\s}\neq
F^\pm_{\s\n} \ee 

The corresponding scalars read \bea
&F^+\equiv g^{\n\s} F^+_{\n\s}=g^{\n\s}\left(\nabla_\l A^\l_{\n\s}-\nabla_\s A^\l_{\n\l}\right)= g^{\n\s}\nabla_\l A^\l_{\n\s}-\nabla^\l A_\l\nonumber\\
&F^-\equiv F^{-\l}\,_\l=\nabla^\l A_\l-g^{\n\s}\nabla_\l
A^\l_{\n\s}=-F^+ \nonumber\\
&O^+\equiv g^{\n\s}O^{+}_{\n\s}=A_{\a}A^\a-g^{\n\s}A^\l_{\a\s}A^\a_{\n\l}\nonumber\\
&O^-\equiv O^{-\l}_{\quad\l}=g^{\n\r}A^\a_{\l\r}A^\a_{\n\l}-A_{\a}A^\a=-O^+
\eea  

\par
Now, we take our action (\ref{Fl}).

It is clear that when expanding around these background fields, i.e. when the connection is the Levi-Civita one, there are many relationships with the preceding operators, to wit
\bea
&\mathcal{O}_1=2\mathcal{O}_2=-\mathcal{O}_3=2\mathcal{O}_4=\mathcal{O}_5\nonumber\\
&\mathcal{O}_6=\mathcal{O}_7=\mathcal{O}_{8}=\mathcal{O}_{9}=-\mathcal{O}_{10}=-\mathcal{O}_{11} \eea
The twelve constants collapse to only three:
\bea&\alpha_1=g_1+\frac{1}{2}g_2-g_3+\frac{1}{2}g_4+g_5\nonumber\\
&\alpha_2=g_6+g_7+g_{8}+g_{9}-g_{10}-g_{11}\nonumber\\
&\alpha_3=g_{12}\eea
so the lowest order in the expansion of the action reduces to
\be\label{lo} 
S_0=\int d^nx
\sqrt{|\bg|}\left(\alpha_1~\widebar{R}_{\m\n\r\s}~\widebar{R}^{\m\n\r\s}+\alpha_2~\widebar{R}_{\m\n}~\widebar{R}^{\m\n}+\alpha_3~\widebar{R}^2\right)\ee

The equations of motion are given by the vanishing of the tadpoles. For the metric, this reads

\bea \left.\delta S\right|_{g_{\m\n}=\bg_{\m\n}}&=\int
d^n x~\kappa\sqrt{|\bg|}\left\{\frac{1}{2}\widebar{g}^{\alpha\beta}\widebar{\cL}
-2\a_1\widebar{R}_{\mu\nu\rho}^{~~~~\a}\widebar{R}^{\mu\nu\rho\b}
-2q_1\widebar{R}_{\m}^{~\alpha}\widebar{R}^{\m\b}-\right.\nonumber\\
&\left.-2q_2\widebar{R}^{\mu\alpha\beta\n}\widebar{R}_{\mu\n}
-2\a_3 \widebar{R}^{\alpha\beta}\widebar{R}\right\} \delta
g_{\alpha\beta}\eea
where we define two more combinations of $g$ constants
\bea
&q_1=g_6+g_7-\frac{1}{2}g_{10}-\frac{1}{2}g_{11}\nonumber\\
&q_2=-g_{8}-g_{9}+\frac{1}{2}g_{10}+\frac{1}{2}g_{11}
\eea
We have relegated most general formulas to the Appendix \ref{A}.
It is immediate to check that the EM are identically satisfied for any Riemannian maximally symmetric, constant curvature manifold, where
\be 
\bR_{\a\b\g\d}=-{2 \l\over
	(n-1)(n-2)}\left(\bg_{\a\g}\bg_{\b\d}-\bg_{\a\d}\bg_{\b\g}\right)=\pm\dfrac{1}{L^2}\left(\bg_{\a\g}\bg_{\b\d}-\bg_{\a\d}\bg_{\b\g}\right)
\ee
With our conventions, the scalar curvature is related to the cosmological constant through
\be 
R=-{2 n\over n-2}\l 
\ee
This means that a priori both de Sitter (positive cosmological constant, but negative curvature) and anti de Sitter (negative cosmological constant and positive curvature) are possible vacua for our quadratic theories. In the following expansions we have restricted ourselves to negative values for the cosmological constant (which is the sphere $S_n$ with our conventions) for definiteness. The problem to find more solutions to the above equations is of course an interesting although daunting task.

On the other hand, the connection tadpole reads
	\bea \left.\delta S\right|_{g_{\mu\nu}=\bar{g}_{\mu\nu}}&=\int
d^n x~\sqrt{|\bg|}\left\{
4\a_1\bar{R}_{\lambda}^{~\alpha\rho\beta}\bar{\nabla}_\rho+2q_1\left(\bar{R}^{\alpha\beta}\bar{\nabla}_\lambda
-\delta^{\beta}_{\lambda}\bar{R}^{\alpha\m}\bar{\nabla}_\m\right)+\right.\nonumber\\
&\left.+2q_2\left(\bar{R}_{\lambda}^{~\beta}\bar{\nabla}^\alpha
-\bar{g}^{\alpha\beta}\bar{R}_{\lambda}^{~\m}\bar{\nabla}_\m
\right)+2\a_3\left(\bar{g}^{\alpha\beta}\bar{R}\bar{\nabla}_\lambda-\delta^{\beta}_{\lambda}\bar{R}\bar{\nabla}^\alpha\right)\right\}\delta
\Gamma^{\lambda}_{\alpha\beta}
\eea
The quadratic term in the expansion can be written as
\be S^{(2)}=\int \sqrt{|\bg|}d^n x~
\bigg\{
\dfrac{1}{2}h_{\m\n}M^{\m\n\r\s}h_{\r\s}+h_{\m\n}N^{\m\n\r\s}_\l
A^\l_{\r\s}+\dfrac{1}{2}A^{\a}_{\m\n}~K^{\m\n\r\s}_{\a\b}~A^\b_{\r\s}\bigg\}\label{lagrangian}
\ee

Here the total mass dimension of the operators reads as follows. 
\be
\left[M\right]=2
\ee
But $M$ is proportional to $g \kappa^2$ (where $g$ is a generic coupling constant) times some background squared; there is then no room for derivatives (momenta) 
\be
M\sim \frac{g~\kappa^2}{L^4}
\ee
where we have assumed that the background curvature is $\sim L^{-2}$.
On the other hand,
\be
\left[N\right]=\frac{n}{2}
\ee
$N$ is proportional to $g \kappa$; so that the rest has mass dimension 3, namely one background field plus one momentum. That is
\be
N\sim 
\frac{g~  \kappa~ ~ k}{L^2}
\ee
Finally,
\be
\left[K\right]=n-2
\ee
This is proportional to $g$ only; so that the rest has dimension 2. There are terms with one background, and also terms with two derivatives. In the ultraviolet ($k L >> 1,\quad k\to\infty$) 
\be
K\sim g ~k^2
\ee To be specific, the different operators appearing are 
\begin{align}
M^{\a\b\g\e}&=\kappa^2\left\{\left(\frac{1}{4}\widebar{g}^{\a\b}\widebar{g}^{\g\e}-\frac{1}{2}\widebar{g}^{\alpha\gamma}\widebar{g}^{\beta\epsilon}\right)
\left(\alpha_1~\widebar{R}_{\m\n\r\s}~\widebar{R}^{\m\n\r\s}+\alpha_2~\widebar{R}_{\m\n}~\widebar{R}^{\m\n}+\alpha_3~\widebar{R}^2\right)+\right.
\nonumber\\
&\left.-\widebar{g}^{\g\e}\left(2\a_1\widebar{R}_{\mu\nu\rho}^{~~~~\a}\widebar{R}^{\mu\nu\rho\b}
+2q_1\widebar{R}_{\m}^{~\alpha}\widebar{R}^{\m\b}+2q_2\widebar{R}^{\mu\alpha\beta\n}\widebar{R}_{\mu\n}
+2\a_3 \widebar{R}^{\alpha\beta}\widebar{R}\right)+\right.\nonumber\\
&\left.+2\left(2\a_1+g_1+\frac{1}{2}g_2\right)\widebar{R}^{~~~~\a}_{\mu\n\r}\widebar{R}^{\mu\n\r\g}\widebar{g}^{\beta\epsilon}
+2\left(2q_1+g_8\right)\widebar{R}^{~\a}_{\mu}\widebar{R}^{\mu\gamma}\widebar{g}^{\beta\epsilon}
+4q_2\widebar{R}^{\mu\alpha\gamma\n}\widebar{R}_{\mu\n}\widebar{g}^{\beta\epsilon}+\right.\nonumber\\
&
\left.+4\a_3\widebar{R}^{\alpha\gamma}\widebar{R}\widebar{g}^{\beta\epsilon}+2\left(g_8+g_{9}\right)\widebar{R}_{\mu~~\n}^{~\a\b}\widebar{R}^{\mu\g\e\n}
-2g_3\widebar{R}_{\mu\n}^{~~~\a\gamma}\widebar{R}^{\m\n\beta\epsilon}
+2g_4\widebar{R}_{\mu~\n}^{~\a~\g}\widebar{R}^{\n\b\m\e}+\right.\nonumber\\
&
\left.+2g_5 \widebar{R}_{\mu~\n}^{~\a~\g}\widebar{R}^{\n\e\m\b}+ 2\left(g_{10}+g_{11}\right)
\widebar{R}_{\m}^{~\g}\widebar{R}^{\m\alpha\beta\epsilon}
+2\left(g_6+g_7-g_{8}\right)\widebar{R}^{\alpha\gamma}\widebar{R}^{\beta\epsilon}
+2g_{12}\widebar{R}^{\alpha\beta}\widebar{R}^{\gamma\epsilon}\right\}+\nonumber\\
&+\{\a\leftrightarrow\b\}+\{\g\leftrightarrow\e\}+\{\a\b\leftrightarrow \g\e\}
\end{align}where $\{\a\leftrightarrow\b\}+\{\g\leftrightarrow\e\}$ stands for the symmetrization under the exchange of $\a,\b$ and $\g,\e$ respectively and  $\{\a\b\leftrightarrow \g\e\}$ refers to the symmetrization under the interchange of $\a,\b$ and $\g,\e$.

The mixed graviton-connection piece reads, still in a somewhat symbolic way, where we indicate explicitly the graviton, whereas the connection is implicit
\vspace{-1cm}
\bea
N_{\gamma\epsilon~~\lambda}^{\quad\alpha\beta}A^{\lambda}_{\alpha\beta}&=\kappa\bg_{\g\e}\sum_{I=1}^{I=12}g_I
R^{\mu}_{\quad\nu\rho\sigma}(D_I)_{\mu\mu'}^{\nu\nu'\rho\rho'\sigma\sigma'}\d^{\m'}_\l\d^\a_{\n'}(\d^\b_{\s'}\bn_{\r'}-\d^\b_{\r'}\bn_{\s'})A^{\lambda}_{\alpha\beta}\nonumber +\\
&+2g_1\kappa\left\{
\widebar{R}_{\g\nu\rho\sigma}F_{\e}^{\quad\nu\rho\sigma}
-\widebar{R}_{\mu\g\rho\sigma}F_{\quad\e}^{\mu\quad\rho\sigma}
-\widebar{R}_{\mu\nu\g\sigma}F_{\quad\e}^{\mu\nu\quad\sigma}
-\widebar{R}_{\mu\nu\rho\g}F_{\quad\quad\e}^{\mu\nu\rho}\right\}+\nonumber\\
&+2g_2\kappa\left\{
\widebar{R}_{\g\nu\rho\sigma}F_{\e}^{\quad\rho\nu\sigma}
-\widebar{R}_{\mu\g\rho\sigma}F_{\quad\e}^{\mu\rho\quad\sigma}
-\widebar{R}_{\mu\nu\g\sigma}F_{\quad\e}^{\mu\quad\nu\sigma}
-\widebar{R}_{\mu\nu\rho\g}F_{\quad\quad\e}^{\mu\rho\nu}\right\}+\nonumber\\
&+2g_3\kappa\left\{
-\widebar{R}_{\mu\nu\g\sigma}F_{\quad\e}^{\nu\mu\quad\sigma}
-\widebar{R}_{\mu\nu\rho\g}F_{\quad\e}^{\nu\mu\rho\quad}\right\}+\nonumber\\
&+2g_4\kappa\left\{
-\widebar{R}_{\mu\g\rho\sigma}F_{\quad\e}^{\rho\quad\mu\sigma}
-\widebar{R}_{\mu\nu\rho\g}F_{\quad\e}^{\rho\nu\mu}\right\}+\nonumber\\
&
+2g_5\kappa\left\{
-\widebar{R}_{\mu\g\rho\sigma}F_{\quad\e}^{\rho\sigma\mu}
-\widebar{R}_{\mu\nu\rho\g}F_{\quad\e}^{\rho\quad\mu\nu}\right\}+\nonumber\\
&+2g_6\kappa\left\{
-\widebar{R}^{+}_{\g\sigma}F^{+\sigma}_{\e}
-\widebar{R}^{+}_{\nu\g}F_{+\e}^{\nu}\right\}
+2g_7\kappa\left\{
-\widebar{R}^{+}_{\g\sigma}F_{+\e}^{\sigma}
-\widebar{R}^{+}_{\sigma\g}F_{\e}^{+\sigma}\right\}+\nonumber\\
&+2g_{8}\kappa\left\{
\widebar{R}^{-}_{\g\sigma}F_{\e}^{-\sigma}
-\widebar{R}_{\mu\g\e\sigma}F_{-}^{\mu\sigma}
-\widebar{R}^{-}_{\mu\sigma}F_{\quad\g\e}^{\mu\quad\quad\sigma}
-\widebar{R}^{-}_{\mu\g}F_{-\e}^{\mu}\right\}+\nonumber\\
&+2g_{9}\kappa\left\{
-\widebar{R}_{\rho\g\e\sigma}F_{-}^{\sigma\rho}
-\widebar{R}_{-}^{\mu\sigma}F_{\sigma\g\e\mu}\right\}+\nonumber\\
&+g_{10}\kappa\left\{
-\widebar{R}_{+}^{\nu\sigma}F_{\nu\g\e\sigma}
-F_{+}^{\nu\sigma}\widebar{R}_{\nu\g\e\sigma}
-\widebar{R}^{+}_{\nu\g}F^{\nu}_{-\e}
-F^{+}_{\nu\g}\widebar{R}^{\nu}_{-\e}\right\}+\nonumber\\
&+g_{11}\kappa\left\{
-\widebar{R}_{+}^{\nu\sigma}F_{\sigma\g\e\nu}
-F_{+}^{\nu\sigma}\widebar{R}_{\sigma\g\e\nu}
-\widebar{R}^{+\sigma}_{\g}F^{-}_{\sigma\e}
-F^{+\sigma}_{\g}\widebar{R}^{-}_{\sigma\e}\right\}+\nonumber\\
&+2g_{12}\kappa\left\{ -\widebar{R}^{+}_{\g\e}F
-\widebar{R}F^{+}_{\g\e}\right\} +\{\a\leftrightarrow\b\}+\{\g\leftrightarrow\e\}
\eea
Finally, the operator relating the connection fluctuations reads
\bea
K_{\lambda\quad\tau}^{\alpha\beta\gamma\epsilon}
&=\left[8\a_1\delta^\alpha_\tau\bR_\lambda^{\quad\gamma\beta\epsilon}
+4q_1\delta^\alpha_\tau\left(\delta^\beta_\lambda\bR^{\gamma\epsilon}
-\delta^\epsilon_\lambda\bR^{\gamma\beta} \right)+\right.\nonumber\\
&\left.+4q_2\delta^\alpha_\tau\left(\bg^{\gamma\beta}\bR_\lambda^{\quad\epsilon}
-\bg^{\gamma\epsilon}\bR_\lambda^{\quad\beta}\right)
+4\a_3\delta^\alpha_\tau\left(\delta^\beta_\lambda\bg^{\gamma\epsilon}\bR
-\delta^\epsilon_\lambda\bg^{\gamma\beta}\bR \right)\right]
+\nonumber\\
&+2\widebar{\nabla}^\e\widebar{\nabla}^{\b}\left[\bg_{\l\tau}\bg^{\a\g}\left(2g_1+g_2-g_8\right)
+\d^{\g}_{\l}\d^{\a}_{\t}\left(2g_3-g_5-g_{9}\right)+\right.\nonumber\\
&\left.+\d^{\a}_{\l}\d^{\g}_{\t}\left(-g_7-g_{12}\right)
\right]+\nonumber\\
&+
2\widebar{\nabla}_\t\widebar{\nabla}^{\b}\left[\bg^{\a\g}\d^{\e}_{\l}\left(2g_4+2g_5-g_{10}-g_{11}\right)
+\d^{\a}_{\l}\bg^{\g\e}\left(-g_{11}+2g_{12}\right)\right]+\nonumber\\
&+2\widebar{\nabla}_\t\widebar{\nabla}_{\l}\left[-\bg^{\a\g}\bg^{\b\e}\left(g_4+g_5
+g_6+g_7 \right)
-\bg^{\a\b}\bg^{\g\e}\left(g_{9}+g_{12}\right)\right]+\nonumber\\
&+2\widebar{\nabla}_\t\widebar{\nabla}^{\e}\left[\bg^{\g\a}\d^{\b}_{\l}\left(2g_6+2g_7+g_{11}\right)
+\bg^{\a\b}\d^{\g}_{\l}\left(2g_{9}+g_{10}+g_{11}\right)
\right]+\nonumber\\
&+2\widebar{\nabla}^\a\widebar{\nabla}^{\b}\left[2\bg_{\l\t}\bg^{\g\e}g_8
+\d^{\e}_{\t}\d^{\g}_{\l}g_{10}
\right]+\nonumber\\
&+2A^{\l}_{\a\b}\Box
A^{\t}_{\g\e}\left[\bg_{\l\tau}\bg^{\a\g}\bg^{\b\e}\left(-2g_1-g_2\right)
-\bg_{\l\tau}\bg^{\a\b}\bg^{\g\e}g_8
+\bg^{\a\g}\d^{\e}_{\l}\d^{\b}_{\t}\left(-2g_3-g_4\right)-\right.\nonumber\\
&\left.
-\bg^{\a\g}\d^{\e}_{\t}\d^{\b}_{\l}g_6
-\d^{\g}_{\t}\d^{\e}_{\l}\bg^{\a\b}g_{10}
\right]+\nonumber\\
&+\{\a\leftrightarrow\b\}+\{\g\leftrightarrow\e\}+\{\l\a\b\leftrightarrow\t\g\e\}
\eea
 with $\{\a\leftrightarrow\b\}+\{\g\leftrightarrow\e\}$ defined before and $\{\l\a\b\leftrightarrow\t\g\e\}$ referring to the symmetrization under the interchange of $\l,\a,\b$ and $\t,\g,\e$.

\section{Interaction between external sources.}
We have already mentioned the enormity of the theory space  we have been considering. Our main interest, however, is to find a theory describing the gravitational interaction.
Let us now discuss our general strategy in order to determine the correct
physical effect of external sources. What we want is to characterize  the  physical sources that interact gravitationally in our theory.

To begin with, assume that we introduce two external sources of dimension $\left[T\right]=1+{n\over 2}$, one coupled to the graviton
\be
\int d(vol)~T_{\m\n} h^{\m\n}
\ee
In order for this term to be gauge invariant under linearized gauge transformations
\be
\d h_{\m\n}=\bn_\m\xi_\n+\bn_\n\xi_\m
\ee
The source needs to be symmetric and background-covariantly conserved
\bea
&T_{\m\n}=T_{\n\m}\nonumber\\
&\bn_\m T^{\m\n}=0
\eea

\par
We could  also introduce another  source  coupled to the connection with dimension $[J]=n-1$
\be
\int d(vol)~J^{\m\n\l}~A_{\m\n\l}
\ee
where the source is got to be symmetric in the last two indices $
J_{\a\b\g}=J_{\a\g\b}$. 
Gauge invariance now means that
\be
\pounds(\xi)\,J_{\a\b\g}=0
\ee

Let us think about the relationship between the response to a graviton source $T_{\m\n}$ which we denote by $h_{\m\n}$ and the response to a connection source $J_{\a\b\g}$ which we denote by $A_{\a\b\g}$.
\par
In GR the graviton fluctuation is given by
\be
h_{\m\n}=\int \Delta^{GR}_{\m\n}\,^{\r\s}~T_{\r\s}
\ee

If the connection were Levi-Civita and the theory were formulated in second order, then the relationship between the responses to both sources reads
\bea
&A_{\m\n\r}=-{1\over 2}\bigg\{-\nabla_\m h_{\n\r}+\nabla_\n h_{\m\r}+\nabla_\r h_{\m\n}\bigg\}=\nonumber\\
&=-{1\over 2}\int d^n y~\bigg\{-\nabla_\m \Delta^{GR}_{\n\r}\,^{\a\b}(x,y)+\nabla_\n \Delta^{GR}_{\m\r}\,^{\a\b}(x,y)+\nabla_\r \Delta^{GR}_{\m\n}(x,y)\bigg\}~T_{\a\b}(y)
\eea
We expect that  this is related in some limit to
\be
\langle A_{\m\n\r}\rangle\equiv {\d W[J]\over \d J^{\m\n\r}}= K^{-1}_{\m\n\r}\,^{\a\b\g} J_{\a\b\g}
\ee
In this limit 
\be
J_{\a\b\g}=-{1\over 2}~\int K_{\a\b\g}\,^{\m\n\r}\int d^n y~\bigg\{-\nabla_\m \Delta^{GR}_{\n\r}\,^{\a\b}(x,y)+\nabla_\n \Delta^{GR}_{\m\r}\,^{\a\b}(x,y)+\nabla_\r \Delta^{GR}_{\m\n}(x,y)\bigg\}T_{\a\b}(y)
\ee
Then we should recover the GR result for the free energy (at least in the lowest order approximation) , namely

\be
W[T]=C\int {d^4 k\over k^2}~\left(|T_{\m\n}(k) T^{\m\n}(k)|-{1\over 2} |T(k)|^2\right)
\ee

This presumably yields a general idea of what is what we should expect in the first order case.

The gaussian path integral yields for the free energy the result (up to an additive constant)

\bea
& W\left[J_{\a\b\g},T_{\m\n}\right]\equiv - \log~Z\left[J_{\a\b\g},T_{\m\n}\right]=\int ~\sqrt{\bg}~d^n x~\bigg\{-{1\over 2} T_{\m\n} \left(M^{-1}\right)^{\m\n\r\s} T_{\r\s} + \nonumber\\
&- {1\over 2}\left(N_{\m\n\l}^{\a\b}~
\left(M^{-1}\right)_{\a\b}^{\r\s}~ T_{\r\s} - J_{\m\n\l} \right)
\left( K^{\m\n\l a b c}- N^{\m\n\l}_{u v} \left(M^{-1}\right)^{u v w x} N_{wx}^{ a b c}\right)^{-1}~\nonumber\\
&\left(N_{a b c}^{u v}\left( M^{-1}\right)_{u v w x } T^{w x} -
J_{a b c}\right) \bigg\} \label{freeenergy}
\eea
Please note that the strength of the interaction between external graviton  sources  is always a contact one
\be
\langle T T \rangle\sim \frac{L^4 }{g \kappa^2}
\ee
The mixing between the graviton source and the connection source, on the other hand, is  ultralocal 
\be
\langle T J\rangle\sim {L^2 \over g \kappa k}
\ee
Finally, the interaction between connection sources allows a long-range potential
\be
\langle J J\rangle\sim {1\over g k^2}
\ee
\subsection{Flat background.}

Let us now work out in turn with some detail the structure of the fluctuations around a flat  background (it is {\em not} then a background gauge calculation, which should be background independent). 
Assume then
 \be \bg_{\m\n}=\eta_{\m\n} \ee
  so that  the whole
contribution to $M_{\m\n\r\s}$ comes from the gauge fixing term (in case we choose to gauge fix the graviton piece).
This has the following problem. We have only four gauge
parameters, whereas there are ten components in the graviton
field. The mismatch means that there are undamped components in
the graviton field. For example, with our gauge fixing, only the
graviton trace, $h$ gets a kinetic term quadratic in derivatives.
Besides, the mixing graviton/connection also vanishes in this
background, $\widebar{N}=0$. Defining the traceless component of the
graviton field \be h^T_{\m\n}\equiv h_{\m\n}-{1\over
	n}~h~\eta_{\m\n} \ee the path integral over ${\cal D} h^T_{\m\n}$ is
not bounded (it would put restrictions on the source,
$\d\left(T^T_{\m\n}\right)$), and so is the total functional
integral.
 
\par
This should be contrasted with what was derived earlier for the Einstein-Hilbert case, where the integration over graviton fluctuations yields a delta-function that defines the connection $\widebar{A}$ in terms of the external source. The main difference is that in the Einstein-Hilbert case the off-diagonal graviton-gauge term $h \widebar{N} A$ did not vanish when in a flat background, so that the path integral could be interpreted as a Dirac delta by analytic continuation. Here what happens is that this same term $\widebar{N}$ does vanish in a flat background.
\par
It is however still possible to define the theory in Minkowski space assuming that gravitation is defined by the three-index field $A_{\m\n\l}$ exclusively and normalizing the path integral accordingly, {\em id est}
\be\label{Carmel}\displaystyle
e^{iW}=\dfrac{Z[J_{\m\n\l}]}{Z[0]}\equiv {\displaystyle\int {\cal D} h_{\m\n}~{\cal D} A_{\m\n\l}~e^{i S\left[ h_{\a\b}, A_{\m\n\l}; J_{\a\b\g}\right]} \over \displaystyle\int  {\cal D} h_{\m\n}~e^{i S\left[ h_{\a\b}, A_{\m\n\l}\right]}}
\ee
This is more or less equivalent to consider that all the graviton dynamics is to be obtained as a consequence of the dynamics of the three-index field $A_{\m\n\l}$, considered as a composite field of sorts. 
\par
 
\par
In this case we can easily invert the $K_{\a_1 \b_1 \g_1}\,^{\a_2 \b_2 \g_2}$ operator by imposing

\be K^{\l\a\b}_{\m\n\s}(K^{-1})^{\m\n\s}_{\t\g\e}=\dfrac{1}{2}\left( \delta^{\alpha}{}_{\epsilon} \delta^{\beta}{}_{\gamma} \delta^{\lambda}{}_{\tau} +  \delta^{\alpha}{}_{\gamma} \delta^{\beta}{}_{\epsilon} \delta^{\lambda}{}_{\tau}\right)\ee

although the answer is a bit cumbersome, namely

\bea (K^{-1})^{\lambda~~\tau}_{\alpha\beta\gamma\epsilon}&=\dfrac{1}{k^2}\Big(\beta_1 \eta_{\alpha \beta} \eta_{\gamma \epsilon} \eta^{\lambda \tau} + \beta_2 \eta_{\alpha \gamma} \eta_{\beta \epsilon} \eta^{\lambda \tau}  + \beta_3 \delta_{\alpha}{}^{\lambda} \delta_{\gamma}{}^{\tau} \eta_{\beta \epsilon} + \beta_4 \delta_{\alpha}{}^{\lambda} \delta_{\beta}{}^{\tau} \eta_{\gamma \epsilon}+\nonumber\\&+\beta_5 \delta_{\alpha}{}^{\tau} \delta_{\gamma}{}^{\lambda} \eta_{\beta \epsilon}+ \beta_6 \eta_{\alpha \beta} \eta_{\gamma \epsilon} \dfrac{k^{\lambda} k^{\tau}}{k^2} + \beta_7 \eta_{\alpha \gamma} \eta_{\beta \epsilon} \dfrac{k^{\lambda} k^{\tau}}{k^2} + \beta_8 \delta_{\beta}{}^{\tau} \eta_{\gamma \epsilon} \dfrac{k_{\alpha} k^{\lambda}}{k^2}  +\nonumber\\&+ \beta_9 \delta_{\gamma}{}^{\tau} \eta_{\beta \epsilon} \dfrac{k_{\alpha} k^{\lambda}}{k^2}   + \beta_{10} \delta_{\alpha}{}^{\tau} \eta_{\beta \epsilon} \dfrac{k_{\gamma} k^{\lambda}}{k^2} + \beta_{11} \delta_{\epsilon}{}^{\tau} \eta_{\alpha \beta} \dfrac{k_{\gamma} k^{\lambda}}{k^2}  +  \beta_{12} \eta_{\gamma \epsilon} \eta^{\lambda \tau} \dfrac{k_{\alpha} k_{\beta}}{k^2} +\nonumber\\&
+  \beta_{13} \delta_{\gamma}{}^{\lambda} \delta_{\epsilon}{}^{\tau} \dfrac{k_{\alpha} k_{\beta}}{k^2} + \beta_{14} \eta_{\beta \epsilon} \eta^{\lambda \tau} \dfrac{k_{\alpha} k_{\gamma}}{k^2} + \beta_{15} \delta_{\beta}{}^{\lambda} \delta_{\epsilon}{}^{\tau} \dfrac{k_{\alpha} k_{\gamma}}{k^2}   + \beta_{16} \delta_{\beta}{}^{\tau} \delta_{\epsilon}{}^{\lambda} \dfrac{k_{\alpha} k_{\gamma}}{k^2} +\nonumber\\&+ \beta_{17} \eta_{\gamma \epsilon} \dfrac{k_{\alpha} k_{\beta} k^{\lambda} k^{\tau}}{k^4}  + \beta_{18} \eta_{\beta \epsilon} \dfrac{k_{\alpha} k_{\gamma} k^{\lambda} k^{\tau}}{k^4}   + \beta_{19} \delta_{\epsilon}{}^{\tau} \dfrac{k_{\alpha} k_{\beta} k_{\gamma} k^{\lambda}}{k^4}  + \beta_{20} \delta_{\beta}{}^{\tau} \dfrac{k_{\alpha} k_{\gamma} k_{\epsilon} k^{\lambda}}{k^4}  +\nonumber\\&   + \beta_{21} \eta^{\lambda \tau} \dfrac{k_{\alpha} k_{\beta} k_{\gamma} k_{\epsilon}}{k^4} + \beta_{22}\dfrac{ k_{\alpha} k_{\beta} k_{\gamma} k_{\epsilon} k^{\lambda}}{k^4} \Big) +\{\a\leftrightarrow\b\}+\{\g\leftrightarrow\e\}+\{\l\a\b\leftrightarrow\t\g\e\}\label{kmenos1}
\eea
where the coefficients $\b_i$ are complicated functions of the coupling constants $g_i$, whose explicit expression is not very illuminating.
What we want in the end is, of course, to recover General Relativity (GR), again, in the lowest order approximation. Therefore, we should be able to predict the Newton potential, plus higher order corrections. This implies that there must be sources $J_{\a\b\g}$ fulfilling that, at the lowest order,

\bea 
&J^{a\b\g}(K^{-1})_{\a\b\g\m\n\r}J^{\m\n\r}=T^{\m\n}\dfrac{1}{2k^2}\left(\eta_{\m\r}\eta_{\n\s}+\eta_{\m\s}\eta_{\n\r}-\eta_{\m\n}\eta_{\r\s}\right)T^{\r\s}=\nonumber\\
&=\frac{T_{ab} T^{ab}}{k^2} -  \frac{T^{a}{}_{a} T^{b}{}_{b}}{2 k^2}\label{sources}
\eea

 Since we are interested in this equality at the lowest order, we can keep only the first five terms in the inverse, since the others will yield corrections to the Newton potential.

Equation \eqref{sources} then reduces to
\bea
&\frac{\beta_1 J_{\a}{}^{\g}{}_{\g} J^{\a\b}{}_{\b}}{k^2} + \frac{\beta_2 J_{\a\b\g} J^{\a\b\g}}{k^2} + \frac{\beta_5 J^{\a\b\g} J_{\b\a\g}}{k^2} + \frac{\beta_4 J^{\a}{}_{\a}{}^{\b} J_{\b}{}^{\g}{}_{\g}}{k^2} + \frac{\beta_3 J^{\a}{}_{\a}{}^{\b} J^{\g}{}_{\b\g}}{k^2}=\nonumber\\
&=\frac{T_{\a\b} T^{\a\b}}{k^2} -  \frac{T^{\a}{}_{\a} T^{\b}{}_{\b}}{2 k^2}
\eea
Assuming, as we did earlier when dealing with the Einstein Hilbert term in the first order formalism, that all physical quantities must be expressed in momentum space in terms of the basic quantities $\eta_{\m\n}$ and $k_\a$, 
\be
J_{\a\b\g}\equiv A k_\a~\eta_{\b\g}+ B\left( k_\b \eta_{\a\g}+k_\g \eta_{\a\b}\right)
\ee
as well as 
\be
T_{\a\b}\equiv t\left(k^2\eta_{\a\b}-k_\a k_\b\right)
\ee
the preceding equation reduces to
\bea
&(\b_1+\b_2)(nA+2B)^2+\b_5\left(A^2+ 2 AB (n+1)+ B^2(n+3)\right)+\nonumber\\
&\quad+(\b_4+\b_3)\left(A+ (n+1)B\right)\left(n A +2 B\right)
= -t^2{(n-1)(n-3)\over 2}
\eea
which has a huge space of solutions.
\par
Alternatively, one may guess a different  ansatz of the type

\be 
J_{\a\b\g}=A j_\a T_{\b\g}+ B (j_b T_{\a\g}+j_\g T_{\a\b})\ee

where $j_\a$ is some conserved vector: $k_\a j^\a=0$. This ansatz illuminates other physical possibilities. In that case the left hand side of \eqref{sources} reads

\bea
&\frac{\bigl(A^2 (\beta_3+\beta_5)+2 A B (2 \beta_2 + \beta_3 + \beta_4 + \beta_5) + B^2 (4 \beta_1 + 2 \beta_2 + \beta_3 + 2 \beta_4 + 3 \beta_5)\bigr) j^{a} j^{b} T_{a}^{c} T_{bc}}{k^2} +\nonumber\\
&+ \frac{j^2 (A^2 \beta_1 + B^2 \beta_3 + A B \beta_4) T^{a}{}_{a} T^{b}{}_{b}}{k^2} +\frac{j^2 \bigl(A^2 \beta_2 + 2 A B \beta_5 + B^2 (2 \beta_2 + \beta_5)\bigr) T_{ab} T^{ab}}{k^2}+\nonumber\\
&+ \frac{\bigl(A^2 \beta_4 + 2 B^2 (\beta_3 + \beta_4) + A B (4 \beta_1 + 2 \beta_3 + \beta_4)\bigr) j^{a} j^{b} T_{ab} T^{c}{}_{c}}{k^2}
\eea

There is no general solution with this ansatz (i.e. without constraints on the $\b_i$). However, if we allow that constraints, we could just set $B=0$ so the previous reduces to

\be
\frac{A^2 j^2 \beta_2 T_{ab} T^{ab}}{k^2} + \frac{A^2 (\beta_3 + \beta_5) j^{a} j^{b} T_{a}{}^{c} T_{bc}}{k^2} + \frac{A^2 j^2 \beta_1 T^{a}{}_{a} T^{b}{}_{b}}{k^2} + \frac{A^2 \beta_4 j^{a} j^{b} T_{ab} T^{c}{}_{c}}{k^2}
\ee

Therefore, by making $\b_4=0$, $\b_3=-\b_5$ and $\b_1=-\dfrac{1}{2}\b_2$ we would achieve the desired result.

This choice, although not unique, proves that the connection sources can be related to the usual ones so we can recover the classical tests of GR again for this ansatz.

\subsection{Curved background.}
It is however possible to assume a constant curvature
background with cosmological constant $\l$ (mass dimension 2). Recall the
behavior of the different operators  in the UV ($k\rightarrow\infty$) 
\bea
&M\sim {g \kappa^2\over L^4}\nonumber\\
&N\sim {g \kappa k\over L^2}\nonumber\\
&K\sim g k^2\nonumber\\
&N~ M^{-1}~N\sim~g k^2
 \eea
  The direct coupling between two
graviton energy-momentum sources is proportional to $M^{-1}\sim
\l^{-2}$, so that it is a contact interaction. The other coupling of
two energy-momentum sources is proportional to
 \be
\left(N~M^{-1}~T\right)\left(K-N~M^{-1}~N\right)^{-1}\left(N~M^{-1}~T\right)\sim
k^0 
\ee
 so that it is again a contact interaction. It is then
unavoidable to introduce connection sources in order to obtain a
non-trivial potential. Indeed the coupling between two such
sources is proportional to 
\be
J\left(K-N~M^{-1}~N\right)^{-1} J\sim{1\over r} 
\ee 
There is some mixing
between the two sources 
\be
\left(N~M^{-1}~T\right)\left(K-N~M^{-1}~N\right)^{-1} J\sim {1\over r^2}
\ee 
At this point it seems that we can dispose of the graviton
source altogether.
\par
To be specific, the graviton EM collapses to

\be \kappa\sqrt{|\bg|}\left(\frac{n}{2}-2\right)\frac{(n-1)}{L^4}\left\{2\a_1+(n-1)\a_2+n(n-1)\a_3\right\}\bg^{\a\b}=0\ee

For $n\neq 4$, this is just a constraint on the coupling constants, that reduces
the number of independent parameters of the most general
lagrangian to eleventh.
\par
{The connection EM  collapses to
 \bea
\int
d^n x~\sqrt{|\bg|}\frac{2}{L^2}\left\{2\a_1+(n-1)\a_2+n(n-1)\a_3\right\} \left(\bg^{\a\b}\bn_\l
-\d^\b_\l\bn^\a\right)\delta
\Gamma^{\lambda}_{\alpha\beta}=0
\eea
which is identically zero because it is a total derivative.

The operator relating the graviton fluctuations in the action \eqref{lagrangian},  reads

\be
M_{\a\b\g\e}=c_1\bg_{\a\b}\bg_{\g\e}+
c_2\left(\bg_{\a\g}\bg_{\b\e}+\bg_{\b\g}\bg_{\a\e}\right)
\ee
where
\bea &c_1=\frac{\kappa^2}{L^4}\left\{2\a_4-(n-1)\left(2-\frac{n}{4}\right)\left[2\a_1+(n-1)\a_2+n(n-1)\a_3\right]\right\}\nonumber\\
&c_2=\frac{\kappa^2}{L^4}\left\{\a_5+(n-1)\left(2-\frac{n}{4}\right)\left[2\a_1+(n-1)\a_2+n(n-1)\a_3\right]\right\}\label{constantsM}\\
&\a_4=-(2g_3-g_5)+(n-2)(g_8+g_{9})-(n-1)(g_{10}+g_{11})+(n-1)^2g_{12}\\
&\a_5=(2g_3+g_8+g_{9})+n(g_4+g_5)+(n-1)(2g_1+g_2+g_{10}+g_{11})+(n-1)^2(g_6+g_7)\nonumber
\eea

while the other two are,

\bea
N^{\a\b}_{\l~~\g\e}&= \frac{\kappa}{L^2}\Bigg( \bigl(-2 g_1 -  ( n-3) g_{9}+ (n-2) g_{10}+ (n-2) g_{11} - (n^2 -3n+2)g_{12}-\nonumber\\ &-  g_2 + 2 g_3  -  g_4 - 2 g_5 + (1-n)g_6  + (1-n)g_7 + (3-n) g_8 \bigr) \d^{\beta }_{\lambda} \bg_{\gamma \epsilon} \bn^{\alpha} +\nonumber\\ &+ (-2 g_{9} + (1-n)g_{10} +(1-n) g_{11}  - 2 g_3 + g_5 ) \d^{\beta}_{ \epsilon} \bg_{\gamma \lambda} \bn^{\alpha} +\nonumber\\ &+ (-2 g_3 + g_5 - 2 g_8) \d^{\beta}_{ \gamma} \bg_{\epsilon \lambda} \bn^{\alpha}  + (2 g_3 - g_5 + 2 g_8) \bg^{\alpha \beta} \bg_{\epsilon \lambda} \bn_{\gamma}  +\nonumber\\ &+ \bigl(8 g_1  +4g_2 - 4 g_3 +4g_4 +6 g_5+4 (n-1)(g_6+g_7)+2(n-1)(g_8+g_9)  +\nonumber\\ &+ (3-2n)(g_{10}+g_{11})+2 n(n-1) g_{12}\bigr) \d^{\alpha}_ {\gamma} \d^{\beta}_{ \lambda} \bn_{\epsilon} +\nonumber\\ &+ \bigl(2 g_{9} + (-1 + n) g_{10}+ (n-1) g_{11} + 2 g_3 -  g_5\bigr) \bg^{\alpha \beta} \bg_{\gamma \lambda} \bn_{\epsilon}+\nonumber\\ &+ \bigl(-8 g_1  -4g_2 + 4 g_3 -4g_4 -6 g_5-4 (n-1)(g_6+g_7)-2(n-1)(g_8+g_9)  -\nonumber\\ &- (3-2n)(g_{10}+g_{11})-2 n(n-1) g_{12}\bigr) \d^{\alpha}_{ \gamma} \d^{\beta}_{ \epsilon} \bn_{\lambda} +\nonumber\\ & \bigl(2 g_1 + (-3 + n) g_{9} + (2-n) g_{10}+ (2-n) g_{11}+ (n^2 -3n+2)g_{12}+g_2 -\nonumber\\ &-2g_3+g_4+2g_5+( n-1) g_6+( n-1) g_7 + (n-3) g_8\bigr) \bg^{\alpha \beta} \bg_{\gamma \epsilon} \bn_{\lambda}\bigg)+\nonumber\\
&+\{\a\leftrightarrow\b\}+\{\g\leftrightarrow\e\}\eea
\bea
K_{\lambda\quad\tau}^{\alpha\beta\gamma\epsilon}
&=
\frac{4}{L^2}\left\{2\a_1+(n-1)\a_2+n(n-1)\a_3\right\}
\left(\bg^{\g\e}\d^\b_\l \d^\a_\t-\bg^{\g\b}\d^\a_\t \d^\e_\l\right)+\nonumber\\
&+2\widebar{\nabla}^\e\widebar{\nabla}^{\b}\left[\bg_{\l\tau}\bg^{\a\g}\left(2g_1+g_2-g_8\right)
+\d^{\g}_{\l}\d^{\a}_{\t}\left(2g_3-g_5-g_{9}\right)+\right.\nonumber\\
&\left.+\d^{\a}_{\l}\d^{\g}_{\t}\left(-g_7-g_{12}\right)
\right]+\nonumber\\
&+
2\widebar{\nabla}_\t\widebar{\nabla}^{\b}\left[\bg^{\a\g}\d^{\e}_{\l}\left(2g_4+2g_5-g_{10}-g_{11}\right)
+\d^{\a}_{\l}\bg^{\g\e}\left(-g_{11}+2g_{12}\right)\right]+\nonumber\\
&+2\widebar{\nabla}_\t\widebar{\nabla}_{\l}\left[-\bg^{\a\g}\bg^{\b\e}\left(g_4+g_5
+g_6+g_7 \right)
-\bg^{\a\b}\bg^{\g\e}\left(g_{9}+g_{12}\right)\right]+\nonumber\\
&+2\widebar{\nabla}_\t\widebar{\nabla}^{\e}\left[\bg^{\g\a}\d^{\b}_{\l}\left(2g_6+2g_7+g_{11}\right)
+\bg^{\a\b}\d^{\g}_{\l}\left(2g_{9}+g_{10}+g_{11}\right)
\right]+\nonumber\\
&+2\widebar{\nabla}^\a\widebar{\nabla}^{\b}\left[2\bg_{\l\t}\bg^{\g\e}g_8
+\d^{\e}_{\t}\d^{\g}_{\l}g_{10}
\right]+\nonumber\\
&+2\Box
\left[\bg_{\l\tau}\bg^{\a\g}\bg^{\b\e}\left(-2g_1-g_2\right)
-\bg_{\l\tau}\bg^{\a\b}\bg^{\g\e}g_8
+\bg^{\a\g}\d^{\e}_{\l}\d^{\b}_{\t}\left(-2g_3-g_4\right)-\right.\nonumber\\
&\left.
-\bg^{\a\g}\d^{\e}_{\t}\d^{\b}_{\l}g_6
-\d^{\g}_{\t}\d^{\e}_{\l}\bg^{\a\b}g_{10}
\right]+ \nonumber\\&+\{\a\leftrightarrow\b\}+\{\g\leftrightarrow\e\}+\{\l\a\b\leftrightarrow\t\g\e\}\eea

Let us work out the zero modes of the operator $M$.

\be
M_{\a\b\g\e}\left(\bn^\g\xi^\e+\bn^\e\xi^\g\right)= 2 c_1 \bg_{\a\b} \bn_\l \xi^\l+2 c_2\left(\bn_\a \xi_\b+\bn_\b\xi_\a\right)=0
\ee

It is easy to see that, by taking the trace, consistency demands that
\be
2 n c_1=-4 c_2
\ee
 The equations of motion put a constraint on the $g_i$ constants of the Lagrangian that implies that this equality is satisfied for every $n$. When this condition is satisfied, the conformal Killing vectors of the manifold are zero modes of $M$. For example, in the case of the sphere $S_n$ the ${(n+1)(n+2)\over 2}$ conformal Killing vectors close the Lie algebra of $O(1,n+1)$.\par

As said before, due to the equations of motion $2nc_1=-4c_2$ and we need to add the gauge fixing term. It is enough to invert the operator to add to the lagrangian

\bea
&L_\text{gf}[h]=-\dfrac{1}{2\rho}\sqrt{|\widebar{g}|}C_\m C^\m\label{gf}
\eea
where
$C_\m C^\m=\bn^\m h_{\m}\bn_\r h^{\r}$. The inverse operator $M^{-1}$ is then obtained by imposing

\be
M_{\a\b\g\e}(M^{-1})^{\g\e\r\s}=\dfrac{1}{2}(\d_\a^\r\d_\b^\s+\d_\a^\s\d_\b^\r)
\ee
This reads

\bea
(M^{-1})^{\a\b\g\e}&=-\dfrac{\Box +  \rho c_1}{4 c_2 \bigl(2\Box + \rho (2 c_1 + \
	c_2)\bigr)}\bg^{\a\b}\bg^{\g\e}+\dfrac{1}{4c_2}\left(\bg^{\a\g}\bg^{\b\e}+\bg^{\b\g}\bg^{\a\e}\right)=\nonumber\\&=\dfrac{\Box +  \rho c_1}{16 c_1\Box }\bg^{\a\b}\bg^{\g\e}-\dfrac{1}{8c_1}\left(\bg^{\a\g}\bg^{\b\e}+\bg^{\b\g}\bg^{\a\e}\right)
\eea

The other term needed to find the free energy \eqref{freeenergy} is

\be
((K- N M^{-1} N)^{-1})_{\a\b~\g\e}^{~\l~~\t}
\ee

This will look as the inverse obtained earlier for $K^{-1}$ in flat space \eqref{kmenos1}, but instead of constants, there will be a set of 22 functions determined by a system of ordinary differential equations similar to the ones solved for simple (but similar) models in the Appendix \ref{ccs}. The explicit expressions for those differential equations is even less illuminating than the flat space expressions so we refrain from considering them further. As an example, and in terms of the arc-length, $s$, and its derivative, $s_\m$,
\bea
((K- N M^{-1} N)&^{-1})_{\a\b~\g\e}^{~\l~~\t}=\beta_1(s) g_{\alpha \beta} g_{\gamma \epsilon} g^{\lambda \tau} + \beta_2(s) g_{\alpha \gamma} g_{\beta \epsilon} g^{\lambda \tau}  + \beta_3(s) \delta_{\alpha}^{\lambda} \delta_{\gamma}^{\tau} g_{\beta \epsilon} + \beta_4(s) \delta_{\alpha}^{\lambda} \delta_{\beta}^{\tau} g_{\gamma \epsilon}+\nonumber\\&+\beta_5(s) \delta_{\alpha}^{\tau} \delta_{\gamma}^{\lambda} g_{\beta \epsilon}+ \beta_6(s) g_{\alpha \beta} g_{\gamma \epsilon} s^{\lambda} s^{\tau} + \beta_7(s) g_{\alpha \gamma} g_{\beta \epsilon} s^{\lambda} s^{\tau}+ \beta_8(s) \delta_{\beta}^{\tau} g_{\gamma \epsilon} s_{\alpha} s^{\lambda} +\nonumber\\&+ \beta_9(s) \delta_{\gamma}^{\tau} g_{\beta \epsilon} s_{\alpha} s^{\lambda}  + \beta_{10}(s) \delta_{\alpha}^{\tau} g_{\beta \epsilon} s_{\gamma} s^{\lambda}+ \beta_{11}(s) \delta_{\epsilon}^{\tau} g_{\alpha \beta} s_{\gamma} s^{\lambda} +  \beta_{12}(s) g_{\gamma \epsilon} g^{\lambda \tau} s_{\alpha} s_{\beta}+\nonumber\\&+  \beta_{13}(s) \delta_{\gamma}^{\lambda} \delta_{\epsilon}^{\tau} s_{\alpha} s_{\beta}+ \beta_{14}(s) g_{\beta \epsilon} g^{\lambda \tau} s_{\alpha} s_{\gamma}+ \beta_{15}(s) \delta_{\beta}^{\lambda} \delta_{\epsilon}^{\tau} s_{\alpha} s_{\gamma}  + \beta_{16}(s) \delta_{\beta}^{\tau} \delta_{\epsilon}^{\lambda} s_{\alpha} s_{\gamma}+\nonumber\\&+ \beta_{17}(s) g_{\gamma \epsilon} s_{\alpha} s_{\beta} s^{\lambda} s^{\tau} + \beta_{18}(s) g_{\beta \epsilon} s_{\alpha} s_{\gamma} s^{\lambda} s^{\tau}  + \beta_{19}(s) \delta_{\epsilon}^{\tau} s_{\alpha} s_{\beta} s_{\gamma} s^{\lambda} +\nonumber\\&+ \beta_{20}(s) \delta_{\beta}^{\tau} s_{\alpha} s_{\gamma} s_{\epsilon} s^{\lambda} + \beta_{21}(s) g^{\lambda \tau} s_{\alpha} s_{\beta} s_{\gamma} s_{\epsilon}+ \beta_{22}(s) s_{\alpha} s_{\beta} s_{\gamma} s_{\epsilon} s^{\lambda} +\nonumber\\&   +\{\a\leftrightarrow\b\}+\{\g\leftrightarrow\e\}+\{\l\a\b\leftrightarrow\t\g\e\}
\eea

\section{Dynamical generation of the Einstein-Hilbert term.}
The theory so far considered is always in the {\em conformal phase}; it is Weyl invariant. This is the symmetry that prevents the appearance of a cosmological constant on the theory and ensures that all counterterms must be inside our list of quadratic operators.

This symmetry is not to be found at low energies, however; which means that it must be broken at some scale. Once this happens, both a cosmological constant and an Einstein-Hilbert term in the lagrangian are not forbidden anymore.
Several scenarios for this breaking can be proposed;
may be the simplest possibility \cite{Salvio}\cite{Einhorn} is through interaction with a minimally coupled  scalar sector
\be
L_s\equiv \sqrt{|g|}\left({1\over 2} g^{\m\n}\pd_\m \phi\pd_\n \phi-V(\phi)\right)
\ee
Quantum corrections will include a term
\be
\Delta L=C_\e R \phi^2
\ee
Were the scalar field to get a nonvanishing vacuum expectation value
\be
\langle\phi\rangle=v
\ee
the counterterm implies an Einstein-Hilbert term
\be
L_{EH}=M^2\sqrt{|g|}R
\ee
The Planck scale $M$ is arbitrary, because it comes about through renormalization; nevertheless the only scale present in the problem to begin with is precisely the symmetry breaking one, $v$.
\section{ Conclusions.}

When considering quadratic in the Riemann tensor gravity theories in the first order formalism, quartic propagators never appear. The ensuing theory naively appears to be both renormalizable and unitary.
\par
In order to laid out the terrain for future work  we have considered all operators with the postulated symmetries and appropriate dimension, with arbitrary coupling constants in front of them. Even if we put some of them equal to zero in the classical lagrangian, quantum effects will generate all the different operators. This makes a grand total of twelve  free coupling constants, which fall naturally into three different groups.
\par
Implicit in this general framework  is that we have to give a physical interpretation not only to the spacetime metric, but also to the connection field (which behaves entirely as a complicated gauge field). It is clear that the theory space is much greater in the first order formulation than in the second order one. One of the first tasks we tackled was to analyze the equations of motion in order to examine what relationship is there between both formulations.
\par
It is precisely this gauge field (id est, the variation of the connection) that encodes all information on the gravitational field.
It is not compulsory to think that there are physical external sources for it, although we have examined this possibility as well. At any rate, we have determined the conditions under which external sources yield a gravitational potential between external energy-momentum sources compatible with the observed one.
\par

\par
The interaction between two external graviton sources has been analyzed both in first-order Einstein-Hilbert and in our quadratic theories. In order for this general approach to be of any physical interest, the theory should generate a mass scale (Newton's constant) through quantum effects. Do not forget that our general framework is Weyl invariant and, correspondingly, all coupling constants are dimensionless as long as the theory remains in the conformal phase. It is then only natural that this process would be related to the spontaneous breaking of Weyl (conformal) symmetry through matter effects, as we suggested earlier in the text, at least in the asymptotically free branch \cite{Stelle} of the theory. Then the Einstein-Hilbert term
\be
S_{EH}= M_p^2\int d^4 x ~\sqrt{|g|}~ g^{\m\n} R_{\m\n}
\ee
which is {\em not} Weyl invariant could be generated by quantum corrections. 
 Were the breaking explicit, it could of course spoil the renormalizability of the theory. But it is known that some theories, like QCD, can dynamically generate a mass scale (like $\Lambda_{QCD}$) while preventing Einstein-Hilbert-like terms to appear in the lagrangian. These terms would then appear in effective low energy theories in terms of different dynamical fields.
 Indeed, in \cite{Holdom} the related conjecture was put forward that the spin 2 ghost that appears in the (second order) quadratic Stelle \cite{Stelle} lagrangians does not appear in the physical spectrum. Similar ideas have been put forward in a related context in \cite{Einhorn}.
\par
If the confining scale of our theory in this sector is $\Lambda_{QG}$, this means that the theory would be strongly coupled in the infrared; but then General Relativity would be an adequate effective theory, playing a somewhat similar relationship with the full theory as  chiral effective theories play with respect to  QCD.
\par

It has been suggested \cite{Dvali} that the ultraviolet completion of some theories involve a mechanism dubbed as {\em classicalization}. The main idea is that instead of a strong coupling phase, the ultraviolet regime involves a high multiplicity of quanta. Owing to this high occupation number, the classical approximation is enough to describe this phase.
\par
These process is suggested by the usual (Schwarzschild) black hole physics and the consequent area law for the entropy. It is not known to what extent they apply to the quadratic in curvature case. There is no Birkhoff theorem that applies there, and there now  three asymptotic families of spherically symmetric solutions \cite{Lu} in the second order formulation.  One of them, that can be matched to an  asymptotically flat solution at spatial infinity without encountering a horizon. Another one that contains both Schwarzschild and non-Schwarzschild black holes. Finally, a third family which is nonsingular and corresponds to vacuum solutions.
\par
These facts shed doubts on whether the {\em classicalization} mechanism would apply to our theory. At any rate, this problem deserves further thought.\par

We have only begun to scratch the surface of this beautiful framework. There remains in particular, to understand the spin content of the three-index gauge field as well as to compute quantum corrections and check explicitly that everything works according to our expectations. This computation is not altogether trivial owing to the appearance of non-minimal operators, which need a special treatment. 
\par
 
 It is plain that this whole approach is related to the age-old question as to what are the fundamental variables in gravitation; the metric or the connection. Work is in progress in this and related matters.

\section*{Acknowledgments}
Two of us (E.A and S.G-M) are grateful to the LBNL and UC Berkeley  for
hospitality in the initial stages of this project. S. G-M. is also grateful to the University of Southampton (U.K.) for their kind hospitality in the final stages of this work. We are grateful to Sergio Hortner, Bert Janssen, C.P. Mart\'in, Tim R. Morris, Raquel Santos-Garc\'ia and Jos Vermaseren for illuminating discussions. Comments by Stanley Deser are always greatly appreciated.
This work has received funding from the European Unions Horizon 2020 research and innovation programme under the Marie Sklodowska-Curie grants agreement No 674896 and No 690575. We also have been partially supported by FPA2012-31880 and FPA2016-78645-P(Spain), COST actions MP1405 (Quantum Structure of Spacetime) and  COST MP1210 (The string theory Universe).  The authors acknowledge the support of the Spanish MINECO {\em Centro de Excelencia Severo Ochoa} Programme under grant  SEV-2012-0249.

\appendix
\newpage
\section{The variation of the Levi Civita is not the Levi-Civita of the variation.}
The fact that there are two different Ricci tensors for a general connection, $R^+_{\m\n}$ and $R^-_{\m\n}$, implies the at first sight surprising fact that it is not the same thing to first put the action on shell (that is, assume the connection is a Levi-Civita one) and then compute its variation or doing things in the opposite order, that is compute the general variation, an then putting the variation on Leci-Civita shell. 

Here we would like to point out that for the Einstein-Hilbert term, this two operations do in fact commmute.
We define the action like
\be S^+_{EH}\equiv\int
d^n x\sqrt{|g|}g^{\m\n}R^+_{\m\n}=\int
d^n x\sqrt{|g|}g^{\m\n}R^\l_{~\m\l\n}\ee
and
\be
 S^-_{EH}\equiv\int
d^n x\sqrt{|g|}g^{\m\n}R^-_{\m\n}=\int
d^n x\sqrt{|g|}g^{\m\n}g^{\r\s}R_{\m\r\s\n}=\int
d^n x\sqrt{|g|}g^{\m\n}R^\l_{~\m\n\l}
\ee

Therefore, doing  Levi-Civita first means that $ S^+_{EH}=-S^-_{EH}$.

On the other hand, if we perform the background field expansion first in $S^-_{EH}$

\bea \left.\delta S^-\right|_{\bg_{\m\n}}&=\int
d^n x~\sqrt{|\bg|}\left\{1+\kappa\frac{h}{2}+\kappa^2\left(\frac{h^2}{8}-\frac{h_{\a\b}h^{\a\b}}{4}\right)\right\}
\left(\bg^{\m\n}-\kappa h^{\m\n}+\kappa^2 h^\m_\l h^{\l\n}\right)\times\nonumber\\
&\times\left\{\pd_\n\left(\bG^\r_{\m\r}+A^\r_{\m\r}\right)-\pd_\r\left(\bG^\r_{\m\n}+A^\r_{\m\n}\right)+\right.
\left(\bG^\r_{\s\n}+A^\r_{\s\n}\right)\left(\bG^\s_{\m\r}+A^\s_{\m\r}\right)-\nonumber\\ &\left.
-\left(\bG^\r_{\s\r}+A^\r_{\s\r}\right)\left(\bG^\s_{\m\n}+A^\s_{\m\n}\right)\right\}=\nonumber\\
&=\bar{S}^-_0+\int
d^n x~\kappa h^{\a\b}\sqrt{|\bg|}\left\{\frac{1}{2}g_{\a\b}\bR^--\bR^-_{\a\b}\right\}-\int
d^n x~\sqrt{|\bg|}\bg^{\m\n}\left\{\delta^{\alpha}_{\mu}
\left(\delta^{\beta}_{\nu}\bar{\nabla}_\lambda-\delta^{\beta}_{\lambda}\bar{\nabla}_\nu\right)\right\}A^{\lambda}_{\alpha\beta}+\nonumber\\
&+\int
d^n x~\kappa^2 h^{\a\b}\sqrt{|\bg|}\left\{\left(\frac{1}{8}\bg_{\a\b}\bg_{\g\e}-\frac{1}{4}\bg_{\a\g}\bg_{\b\e}\right)\bR^-
-\frac{1}{2}\bg_{\a\b}\bR^-_{\g\e}+\bg_{\a\g}\bR^-_{\b\e}\right\}h^{\g\e}-\nonumber\\
&-\int
d^n x~\kappa h^{\g\e}\sqrt{|\bg|}\left\{\left(\frac{1}{2}\bg_{\g\e}\bg^{\m\n}-\d^\m_\g \d^\n_\e\right)\d^\a_\m\left(\delta^{\beta}_{\nu}\bar{\nabla}_\lambda-\delta^{\beta}_{\lambda}\bar{\nabla}_\nu\right)\right\}A^{\lambda}_{\alpha\beta}-\nonumber\\
&-\int
d^n x~\sqrt{|\bg|}\bg^{\m\n}A^{\t}_{\g\e}\left\{\d^\e_\t \d^\g_\l \d^\a_\m \d^\b_\n
-\d^\b_\t \d^\g_\l \d^\a_\m \d^\e_\n\right\}A^{\lambda}_{\alpha\beta}\eea

doing now  Levi-Civita $\bR^+_{\m\n}=-\bR^-_{\m\n}$ y $\bR^+=-\bR^-$ we get

\be \left.\delta S^-\right|_{g_{\m\n}=\bg_{\m\n}}=-\left.\delta S^+\right|_{g_{\m\n}=\bg_{\m\n}}\ee
To conclude, in first order Einstein-Hilbert, these two operations do in fact commute.
\newpage

\section{Details on the background expansion.}\label{A}

The equation of motion for the graviton and the gauge field read respectively
\bea \left.\frac{\delta S}{\delta
	g_{\alpha\beta}}\right|_{g_{\m\n}=\bg_{\m\n}}\hspace{-1em}&=\kappa\sqrt{|\bg|}\Bigg\{\frac{1}{2}\widebar{g}^{\alpha\beta}\widebar{\cL}+g_1\left\{
\widebar{R}^\a_{\quad\nu\rho\sigma}\widebar{R}^{\beta\nu\rho\sigma}
-\widebar{R}^{\quad\a}_{\mu\quad\rho\sigma}\widebar{R}^{\mu\b\rho\sigma}
-\widebar{R}^{\quad\a}_{\mu\nu\quad\sigma}\widebar{R}^{\mu\nu\b\sigma}
-\nonumber\right.\\
&\left.-\widebar{R}^{\quad~\a}_{\mu\nu\rho}\widebar{R}^{\mu\nu\rho\b}\right\}+g_2\left\{
\widebar{R}^\a_{~\nu\rho\sigma}\widebar{R}^{\beta\r\n\sigma}
-\widebar{R}^{~~\a}_{\mu~~\rho\sigma}\widebar{R}^{\mu\r\b\sigma}
-\widebar{R}^{\quad\a}_{\mu\nu~\sigma}\widebar{R}^{\mu\b\n\sigma}
-\widebar{R}^{\quad~\a}_{\mu\nu\rho}\widebar{R}^{\mu\r\n\b}\right\}-\nonumber\\
&-g_3\left\{
\widebar{R}^{\quad\a}_{\mu\nu\quad\sigma}\widebar{R}^{\nu\mu\b\sigma}
+\widebar{R}^{\quad\quad\a}_{\mu\nu\rho}\widebar{R}^{\nu\mu\rho\b}\right\}
-\nonumber\\
&-g_4\left\{
\widebar{R}^{\quad\a}_{\mu\quad\n\sigma}\widebar{R}^{\nu\b\m\sigma}
+\widebar{R}^{\quad\quad\a}_{\mu\nu\rho}\widebar{R}^{\r\n\m\b}\right\}-g_5\left\{
\widebar{R}^{\quad\a}_{\mu\quad\n\sigma}\widebar{R}^{\nu\s\m\b}
+\widebar{R}^{\quad\quad\a}_{\mu\nu\rho}\widebar{R}^{\r\b\m\n}\right\}-\nonumber\\
&-g_6\left\{
\widebar{R}^{\a}_{+\sigma}\widebar{R}^{\b\sigma}_{+}+\widebar{R}^{+\a}_{\nu}\widebar{R}_{+}^{\n\b}\right\}-g_7\left\{
\widebar{R}^{\a}_{+\sigma}\widebar{R}_{+}^{\s\b}+\widebar{R}^{+\a}_{\sigma}\widebar{R}_{+}^{\b\sigma}\right\}-\nonumber\\
&-g_{8}\left\{
-\widebar{R}^{\a}_{-\sigma}\widebar{R}_{-}^{\b\sigma}+\widebar{R}^{\mu\alpha\beta\sigma}\widebar{R}^{-}_{\mu\sigma}
+\widebar{R}^{-}_{\mu\sigma}\widebar{R}^{\mu\a\b\sigma}
+\widebar{R}^{-\a}_{\mu}\widebar{R}_{-}^{\mu\b}\right\}-\nonumber\\
&-g_{9}\left\{
\widebar{R}^{\rho\alpha\beta\sigma}\widebar{R}^{-}_{\sigma\rho}+\widebar{R}^{-}_{\mu\sigma}\widebar{R}^{\sigma\alpha\beta\mu}\right\}-\nonumber\\
&-g_{10}\left\{
\widebar{R}^{+}_{\nu\sigma}\widebar{R}^{\nu\alpha\beta\sigma}+\widebar{R}^{+\a}_{\nu}\widebar{R}^{\nu\b}_{-}\right\}-g_{11}\left\{
\widebar{R}^{+}_{\nu\sigma}\widebar{R}^{\sigma\alpha\beta\nu}
+\widebar{R}^{\a\sigma}_{+}\widebar{R}^{-\b}_{\sigma}\right\}-\nonumber\\
&-g_{12}\left\{
\widebar{R}_{+}^{\alpha\beta}\widebar{R}+\widebar{R}\widebar{R}_{+}^{\alpha\beta}\right\}\Big\}+\{\a\leftrightarrow\b\}\\
 \frac{\delta S}{\delta A^{\lambda}_{\alpha\beta}}\Bigg|_{g_{\mu\nu}=\widebar{g}_{\mu\nu}}&=\sqrt{|\bg|}\Big\{
2g_{1}\left(\widebar{\nabla}_\rho\widebar{R}_{\lambda}^{\quad\alpha\rho\beta}-\widebar{\nabla}_\sigma\widebar{R}_{\lambda}^{\quad\alpha\beta\sigma}\right)+2g_{2}\left(\widebar{\nabla}_\rho\widebar{R}_{\lambda}^{\quad\rho\alpha\beta}-\widebar{\nabla}_\sigma\widebar{R}_{\lambda}^{\quad\beta\alpha\sigma}\right)+\nonumber\\
&+2g_{3}\left(\widebar{\nabla}_\rho\widebar{R}^{\alpha\quad\rho\beta}_{\quad\lambda}-\widebar{\nabla}_\sigma\widebar{R}^{\alpha\quad\beta\sigma}_{\quad\lambda}\right)+\nonumber\\
&+2g_{4}\left(\widebar{\nabla}_\rho\widebar{R}^{\rho\alpha\quad\beta}_{\quad\lambda}-\widebar{\nabla}_\sigma\widebar{R}^{\beta\alpha\quad\sigma}_{\quad\lambda}\right)+2g_{5}\left(\widebar{\nabla}_\rho\widebar{R}^{\rho\beta\quad\alpha}_{\quad\lambda}-\widebar{\nabla}_\sigma\widebar{R}^{\beta\sigma\quad\alpha}_{\quad\lambda}\right)+\nonumber\\
&+2g_{6}\left(\widebar{\nabla}_\lambda\widebar{R}_{+}^{\alpha\beta}-\delta^{\beta}_{\lambda}\widebar{\nabla}_\nu\widebar{R}_{+}^{\alpha\nu}\right)+2g_{7}\left(\widebar{\nabla}_\lambda\widebar{R}_{+}^{\beta\alpha}-\delta^{\beta}_{\lambda}\widebar{\nabla}_\nu\widebar{R}_{+}^{\nu\alpha}\right)+\nonumber\\
&+2g_{8}\left(\widebar{\nabla}^\alpha\widebar{R}_{\lambda}^{-\beta}-\widebar{g}^{\alpha\beta}\widebar{\nabla}_\nu\widebar{R}_{\lambda}^{-\nu}\right)+2g_{9}\left(\widebar{\nabla}^\alpha\widebar{R}_{-\lambda}^{\beta}-\widebar{g}^{\alpha\beta}\widebar{\nabla}_\nu\widebar{R}_{-\lambda}^{\nu}\right)+\nonumber\\
&+g_{10}\left\{\left(\widebar{\nabla}_\lambda\widebar{R}_{-}^{\alpha\beta}-\delta^{\beta}_{\lambda}\widebar{\nabla}_\nu\widebar{R}_{-}^{\alpha\nu}\right)
+\left(\widebar{\nabla}^\alpha \widebar{R}^{+\beta}_{\lambda}-\widebar{g}^{\alpha\beta}\widebar{\nabla}^\nu\widebar{R}^{+}_{\lambda\nu}\right)\right\}+\nonumber\\
&+g_{11}\left\{\left(\widebar{\nabla}_\lambda\widebar{R}_{-}^{\beta\alpha}-\delta^{\beta}_{\lambda}\widebar{\nabla}_\nu\widebar{R}_{-}^{\nu\alpha}\right)
+\left(\widebar{\nabla}^\alpha \widebar{R}^{\beta}_{+\lambda}-\widebar{g}^{\alpha\beta}\widebar{\nabla}^\nu\widebar{R}^{+}_{\nu\lambda}\right)\right\}+\nonumber\\
&+2g_{12}\left(\widebar{g}^{\alpha\beta}\widebar{\nabla}_\lambda\widebar{R}-\delta^{\beta}_{\lambda}\widebar{\nabla}^\alpha\widebar{R}\right)\Bigg\}+\{\a\leftrightarrow\b\}
\eea
The quadratic operator relating graviton-graviton fluctuations is

\bea
M^{\a\b\g\e}&=
\kappa^2\Bigg\{g_{1}\left\{\left(\frac{1}{4}g^{\a\b}g^{\g\e}-\frac{1}{2}g^{\alpha\gamma}g^{\beta\epsilon}\right)R_{\m\n\r\s}~R^{\m\n\r\s}+
+g^{\g\e}\left\{\widebar{R}^\a_{\quad\nu\rho\sigma}\widebar{R}^{\beta\nu\rho\sigma}
-\widebar{R}^{\quad\a}_{\mu\quad\rho\sigma}\widebar{R}^{\mu\b\rho\sigma}
-\right.\right.\nonumber\\
&\left.\left.-\widebar{R}^{\quad\a}_{\mu\nu\quad\sigma}\widebar{R}^{\mu\nu\b\sigma}
-\widebar{R}^{\quad\quad\a}_{\mu\nu\rho}\widebar{R}^{\mu\nu\rho\b}\right\}+2\left\{-\widebar{R}^{\alpha\gamma}_{\quad\rho\sigma}\widebar{R}^{\beta\epsilon\rho\sigma}
-\widebar{R}^{\a\quad\g}_{\quad\nu\quad\sigma}\widebar{R}^{\b\nu\e\sigma}
-\right.\right.\nonumber\\
&\left.\left.-\widebar{R}^{\a\quad~\g}_{\nu\rho}\widebar{R}^{\b\nu\rho\e}
+\widebar{R}^{\quad\a\g}_{\mu\quad\quad\sigma}\widebar{R}^{\mu\b\e\sigma}
+\widebar{R}_{\mu\quad\rho}^{\quad\a\quad\g}\widebar{R}^{\mu\b\rho\e}+\right.\right.\nonumber\\&\left.\left.
+\widebar{R}_{\mu\nu}^{\quad\alpha\gamma}\widebar{R}^{\mu\nu\b\e}
+\widebar{R}^{\quad\a}_{\mu\quad\rho\sigma}\widebar{R}^{\mu\g\rho\sigma}\widebar{g}^{\beta\epsilon}
+\widebar{R}^{\quad\a}_{\mu\nu\quad\sigma}\widebar{R}^{\mu\nu\g\sigma}\widebar{g}^{\beta\epsilon}
+\widebar{R}^{\quad\quad\a}_{\mu\nu\rho}\widebar{R}^{\mu\nu\rho\g}\widebar{g}^{\beta\epsilon}\right\}\right\}+\nonumber\\
&+g_{2}\left\{\left(\frac{1}{4}g^{\a\b}g^{\g\e}-\frac{1}{2}g^{\alpha\gamma}g^{\beta\epsilon}\right)R_{\m\n\r\s}~R^{\m\r\n\s}+
g^{\g\e}\left\{
\widebar{R}^\a_{\quad\nu\rho\sigma}\widebar{R}^{\beta\r\n\sigma}
-\widebar{R}^{\quad\a}_{\mu\quad\rho\sigma}\widebar{R}^{\mu\r\b\sigma}
-\right.\right.\nonumber\\&\left.\left.-\widebar{R}^{\quad\a}_{\mu\nu\quad\sigma}\widebar{R}^{\mu\b\n\sigma}
-\widebar{R}^{\quad\quad\a}_{\mu\nu\rho}\widebar{R}^{\mu\r\n\b}\right\}+2\left\{-\widebar{R}^{\alpha\gamma}_{\quad\rho\sigma}\widebar{R}^{\beta\r\e\sigma}
-\widebar{R}^{\a\quad\g}_{\quad\nu\quad\sigma}\widebar{R}^{\b\e\n\sigma}
-\right.\right.\nonumber\\
&\left.\left.-\widebar{R}^{\a\quad\quad\g}_{\quad\nu\rho}\widebar{R}^{\b\r\n\e}
+\widebar{R}^{\quad\a\g}_{\mu\quad\quad\sigma}\widebar{R}^{\mu\e\b\sigma}
+\widebar{R}_{\mu\quad\rho}^{\quad\a\quad\g}\widebar{R}^{\mu\r\b\e}+\right.\right.\nonumber\\&\left.\left.
+\widebar{R}_{\mu\nu}^{\quad\alpha\gamma}\widebar{R}^{\mu\b\n\e}
+\widebar{R}^{\quad\a}_{\mu\quad\rho\sigma}\widebar{R}^{\mu\r\g\sigma}\widebar{g}^{\beta\epsilon}
+\widebar{R}^{\quad\a}_{\mu\nu\quad\sigma}\widebar{R}^{\mu\g\n\sigma}\widebar{g}^{\beta\epsilon}
+\widebar{R}^{\quad\quad\a}_{\mu\nu\rho}\widebar{R}^{\mu\r\n\g}\widebar{g}^{\beta\epsilon}\right\}\right\}+\nonumber\\
&+g_{3}\left\{\left(\frac{1}{4}g^{\a\b}g^{\g\e}-\frac{1}{2}g^{\alpha\gamma}g^{\beta\epsilon}\right)R_{\m\n\r\s}~R^{\n\m\r\s}
-g^{\g\e}\left\{
\widebar{R}^{\quad\a}_{\mu\nu\quad\sigma}\widebar{R}^{\nu\mu\b\sigma}
+\widebar{R}^{\quad\quad\a}_{\mu\nu\rho}\widebar{R}^{\nu\mu\rho\b}\right\}+\right.\nonumber\\
&\left.+2\left\{
\widebar{R}_{\mu\nu}^{\quad\alpha\gamma}\widebar{R}^{\nu\mu\beta\epsilon}
+\widebar{R}_{\mu\nu\quad\sigma}^{\quad\a}\widebar{R}^{\nu\mu\g\sigma}\widebar{g}^{\beta\epsilon}
+\widebar{R}_{\mu\nu\rho}^{\quad\quad\alpha}\widebar{R}^{\nu\mu\rho\g}\widebar{g}^{\beta\epsilon}\right\}\right\}+\nonumber\\
&+g_{4}\left\{\left(\frac{1}{4}g^{\a\b}g^{\g\e}-\frac{1}{2}g^{\alpha\gamma}g^{\beta\epsilon}\right)R_{\m\n\r\s}~R^{\r\n\m\s}
-g^{\g\e}\left\{
\widebar{R}^{\quad\a}_{\mu\quad\n\sigma}\widebar{R}^{\nu\b\m\sigma}
+\widebar{R}^{\quad\quad\a}_{\mu\nu\rho}\widebar{R}^{\r\n\m\b}\right\}+\right.\nonumber\\
&\left.+2\left\{
\widebar{R}^{\quad\a\quad\g}_{\mu\quad\rho}\widebar{R}^{\rho\b\m\e}
+\widebar{R}^{\quad\a}_{\mu\quad\rho\sigma}\widebar{R}^{\rho\g\m\sigma}\widebar{g}^{\beta\epsilon}
+\widebar{R}_{\mu\nu\rho}^{\quad\quad\alpha}\widebar{R}^{\rho\n\m\g}\widebar{g}^{\beta\epsilon}\right\}\right\}+\nonumber\\
&+g_{5}\left\{\left(\frac{1}{4}g^{\a\b}g^{\g\e}-\frac{1}{2}g^{\alpha\gamma}g^{\beta\epsilon}\right)R_{\m\n\r\s}~R^{\r\s\m\n}
-g^{\g\e}\left\{
\widebar{R}^{\quad\a}_{\mu\quad\n\sigma}\widebar{R}^{\nu\s\m\b}
+\widebar{R}^{\quad\quad\a}_{\mu\nu\rho}\widebar{R}^{\r\b\m\n}\right\}+\right.\nonumber\\
&\left.+2\left\{
\widebar{R}_{\mu\quad\rho}^{\quad\a\quad\gamma}\widebar{R}^{\rho\e\mu\b}
+\widebar{R}^{\quad\a}_{\mu\quad\rho\sigma}\widebar{R}^{\rho\sigma\mu\g}\widebar{g}^{\beta\epsilon}
+\widebar{R}_{\mu\nu\rho}^{\quad\quad\alpha}\widebar{R}^{\rho\g\mu\nu}\widebar{g}^{\beta\epsilon}\right\}\right\}+\nonumber\\
&+g_{6}\left\{\left(\frac{1}{4}g^{\a\b}g^{\g\e}-\frac{1}{2}g^{\alpha\gamma}g^{\beta\epsilon}\right)R^+_{\m\n}~R_+^{\m\n}
-g^{\g\e}\left\{
\widebar{R}^{\a}_{+\sigma}\widebar{R}^{\b\sigma}_{+}+\widebar{R}^{+\a}_{\nu}\widebar{R}_{+}^{\n\b}\right\}+\right.\nonumber\\
&\left.+2\left(\widebar{R}_{+}^{\alpha\gamma}\widebar{R}_{+}^{\beta\epsilon}
+\widebar{R}^{\a}_{+\nu}\widebar{R}^{\g\nu}_{+}\widebar{g}^{\beta\epsilon}
+\widebar{R}^{+\a}_{\mu}\widebar{R}^{\mu\g}_{+}\widebar{g}^{\beta\epsilon}\right)\right\}+\nonumber\\
&+g_{7}\left\{\left(\frac{1}{4}g^{\a\b}g^{\g\e}-\frac{1}{2}g^{\alpha\gamma}g^{\beta\epsilon}\right)R^+_{\m\n}~R_+^{\n\m}
-g^{\g\e}\left\{
\widebar{R}^{\a}_{+\sigma}\widebar{R}_{+}^{\s\b}+\widebar{R}^{+\a}_{\sigma}\widebar{R}_{+}^{\b\sigma}\right\}+\right.\nonumber\\
&\left.+2\left\{ \widebar{R}_{+}^{\alpha\gamma}\widebar{R}_{+}^{\e\b}
+\widebar{R}^{\a}_{+\nu}\widebar{R}^{\n\g}_{+}\widebar{g}^{\beta\epsilon}
+\widebar{R}^{+\a}_{\mu}\widebar{R}^{\g\m}_{+}\widebar{g}^{\beta\epsilon}\right\}\right\}+\nonumber
\eea
\raggedbottom
\bea
&+g_{8}\left\{\left(\frac{1}{4}g^{\a\b}g^{\g\e}-\frac{1}{2}g^{\alpha\gamma}g^{\beta\epsilon}\right)R^-_{\m\n}~R_-^{\m\n}-
\right.\nonumber\\
&\left.-g^{\g\e}\left\{
-\widebar{R}^{\a}_{-\sigma}\widebar{R}_{-}^{\b\sigma}+\widebar{R}^{\mu\alpha\beta\sigma}\widebar{R}^{-}_{\mu\sigma}
+\widebar{R}^{-}_{\mu\sigma}\widebar{R}^{\mu\a\b\sigma}
+\widebar{R}^{-\a}_{\mu}\widebar{R}_{-}^{\mu\b}\right\}+\right.\nonumber\\
&\left.+2\left\{
-\widebar{R}^{\alpha\gamma\epsilon\sigma}\widebar{R}^{\beta}_{-\sigma}
-\widebar{R}^{\a}_{-\sigma}\widebar{R}^{\beta\gamma\epsilon\sigma}
-\widebar{R}_{-}^{\alpha\gamma}\widebar{R}_{-}^{\beta\epsilon}
+\widebar{R}^{\mu\alpha\beta\sigma}\widebar{R}^{\quad\gamma\epsilon}_{\mu\quad\quad\sigma}
+\widebar{R}^{\mu\alpha\beta\gamma}\widebar{R}^{-\epsilon}_{\mu}+\right.\right.\nonumber\\&\left.\left.
+\widebar{R}_{\mu}^{-\gamma}\widebar{R}^{\mu\alpha\beta\epsilon}
+\widebar{R}^{\mu\alpha\gamma\sigma}\widebar{R}^{-}_{\mu\sigma}\widebar{g}^{\beta\epsilon}
+\widebar{R}_{\mu\sigma}^{-}\widebar{R}^{\mu\alpha\gamma\sigma}\widebar{g}^{\beta\epsilon}
+\widebar{R}_{-}^{\mu\alpha}\widebar{R}^{-\gamma}_{\mu}\widebar{g}^{\beta\epsilon}\right\}\right\}+\nonumber\\
&+g_{9}\left\{\left(\frac{1}{4}g^{\a\b}g^{\g\e}-\frac{1}{2}g^{\alpha\gamma}g^{\beta\epsilon}\right)R^-_{\m\n}~R_-^{\n\m}
-g^{\g\e}\left\{
\widebar{R}^{\rho\alpha\beta\sigma}\widebar{R}^{-}_{\sigma\rho}+\widebar{R}^{-}_{\mu\sigma}\widebar{R}^{\sigma\alpha\beta\mu}\right\}+\right.\nonumber\\
&\left.+2\left\{
\widebar{R}^{\mu\alpha\beta\sigma}\widebar{R}^{\quad\gamma\epsilon}_{\sigma\quad\quad\mu}
+\widebar{R}^{\mu\alpha\gamma\sigma}\widebar{R}^{-}_{\sigma\mu}\widebar{g}^{\beta\epsilon}
+\widebar{R}_{-}^{\mu\sigma}\widebar{R}^{\quad\alpha\gamma}_{\sigma\quad\quad\mu}\widebar{g}^{\beta\epsilon}\right\}\right\}+\nonumber\\
&+g_{10}\left\{\left(\frac{1}{4}g^{\a\b}g^{\g\e}-\frac{1}{2}g^{\alpha\gamma}g^{\beta\epsilon}\right)R^+_{\m\n}~R_-^{\m\n}
-g^{\g\e}\left\{
\widebar{R}^{+}_{\nu\sigma}\widebar{R}^{\nu\alpha\beta\sigma}+\widebar{R}^{+\a}_{\nu}\widebar{R}^{\nu\b}_{-}\right\}+\right.\nonumber\\
&\left.+2\left\{
\widebar{R}_{\nu}^{+\gamma}\widebar{R}^{\nu\alpha\beta\epsilon}
+\widebar{R}_{\nu\sigma}^{+}\widebar{R}^{\nu\alpha\gamma\sigma}\widebar{g}^{\beta\epsilon}
+\widebar{R}_{+}^{\nu\alpha}\widebar{R}_{\nu}^{-\gamma}\widebar{g}^{\beta\epsilon}\right\}\right\}+\nonumber\\
&+g_{11}\left\{\left(\frac{1}{4}g^{\a\b}g^{\g\e}-\frac{1}{2}g^{\alpha\gamma}g^{\beta\epsilon}\right)R^+_{\m\n}~R_-^{\n\m}
-g^{\g\e}\left\{
\widebar{R}^{+}_{\nu\sigma}\widebar{R}^{\sigma\alpha\beta\nu}
+\widebar{R}^{\a\sigma}_{+}\widebar{R}^{-\b}_{\sigma}\right\}+\right.\nonumber\\
&\left.+2\left\{
\widebar{R}_{+\sigma}^{\gamma}\widebar{R}^{\sigma\alpha\beta\epsilon}
+\widebar{R}^{+}_{\nu\sigma}\widebar{R}^{\sigma\alpha\gamma\nu}\widebar{g}^{\beta\epsilon}
+\widebar{R}_{+\sigma}^{\alpha}\widebar{R}_{-}^{\sigma\gamma}\widebar{g}^{\beta\epsilon}\right\}\right\}+\nonumber\\
&+g_{12}\left\{\left(\frac{1}{4}g^{\a\b}g^{\g\e}-\frac{1}{2}g^{\alpha\gamma}g^{\beta\epsilon}\right)R^2
-g^{\g\e}\left\{
\widebar{R}_{+}^{\alpha\beta}\widebar{R}+\widebar{R}\widebar{R}_{+}^{\alpha\beta}\right\}\right.\nonumber\\&\left.+2\left\{
\widebar{R}_{+}^{\alpha\beta}\widebar{R}_{+}^{\gamma\epsilon}
+2\widebar{R}_{+}{\alpha\gamma}\widebar{R}\widebar{g}^{\beta\epsilon}
\right\}\right\}\Big\}+\{\a\leftrightarrow\b\}+\{\g\leftrightarrow\e\}
\eea
The mixing term betwen graviton and gauge fluctuations reads
\bea
N^{\a\b}_{\l~~ \g\e}=&\kappa\bg_{\g\e}\sum_{I=1}^{I=12}g_I
R^{\mu}_{\quad\nu\rho\sigma}(D_I)_{\mu\mu'}^{\nu\nu'\rho\rho'\sigma\sigma'}\d^{\m'}_\l\d^\a_{\n'}(\d^\b_{\s'}\bn_{\r'}-\d^\b_{\r'}\bn_{\s'})\nonumber\\
	&\kappa\Bigg\{2g_{1}\left\{\bg_{\gamma\lambda}\bR_{\epsilon}^{\quad\alpha\rho\beta}\widebar{\nabla}_\rho
	-\bg_{\gamma\lambda}\bR_{\epsilon}^{\quad\alpha\beta\rho}\widebar{\nabla}_\rho
	-\delta^{\alpha}_{\gamma}\bR_{\lambda\epsilon}^{\quad\rho\beta}\widebar{\nabla}_\rho
	+\delta^{\alpha}_{\gamma}
	\bR_{\lambda\epsilon}^{\quad\beta\rho}\widebar{\nabla}_\rho
	\right.+\nonumber\\
	&\left.-\bR_{\lambda\quad\epsilon}^{\quad\alpha~\beta}\widebar{\nabla}_\gamma
	+\delta^{\alpha}_{\gamma}
	\bR_{\lambda\quad\epsilon}^{\quad\beta\quad\rho}\widebar{\nabla}_\rho
	-\delta^{\alpha}_{\gamma}
	\bR_{\lambda\quad\quad\epsilon}^{\quad\beta\rho}\widebar{\nabla}_\rho
	+\bR_{\lambda\quad\quad\epsilon}^{\quad\alpha\beta}\widebar{\nabla}_\gamma
	\right\}+\nonumber\\
	&+2g_{2}\left\{\bg_{\gamma\lambda}\bR_{\epsilon}^{\quad\r\a\beta}\widebar{\nabla}_\rho
	-\bg_{\gamma\lambda}\bR_{\epsilon}^{\quad\b\a\rho}\widebar{\nabla}_\rho
	-\delta^{\alpha}_{\gamma}\bR_{\lambda\quad\epsilon}^{\quad\rho\quad\beta}\widebar{\nabla}_\rho
	+\delta^{\alpha}_{\gamma}
	\bR_{\lambda\quad\epsilon}^{\quad\beta\quad\rho}\widebar{\nabla}_\rho
	\right.+\nonumber\\
	&\left.-\bR_{\lambda\epsilon}^{\quad\alpha\beta}\widebar{\nabla}_\gamma
	+\delta^{\alpha}_{\gamma}
	\bR_{\lambda\epsilon}^{\quad\beta\rho}\widebar{\nabla}_\rho
	-\delta^{\alpha}_{\gamma}
	\bR_{\lambda\quad\quad\epsilon}^{\quad\r\b}\widebar{\nabla}_\rho
	+\bR_{\lambda\quad\quad\epsilon}^{\quad\b\a}\widebar{\nabla}_\gamma
	\right\}+\nonumber\\
	&+2g_{3}\left\{-\bR_{\quad\l\epsilon}^{\a\quad\quad\beta}\widebar{\nabla}_\gamma
	+\delta^{\alpha}_{\gamma}
	\bR_{\quad\l\epsilon}^{\b\quad\quad\rho}\widebar{\nabla}_\rho
	-\delta^{\alpha}_{\gamma}
	\bR_{\quad\lambda\quad\epsilon}^{\b\quad\rho}\widebar{\nabla}_\rho
	+\bR_{\quad\lambda\quad\epsilon}^{\a\quad\beta} \widebar{\nabla}_\gamma\right\}+\nonumber\\
	&+2g_{4}\left\{ -\delta^{\alpha}_{\gamma}
	\bR_{\quad\epsilon\lambda}^{\rho\quad\quad\beta}\widebar{\nabla}_\rho
	+\delta^{\alpha}_{\gamma}
	\bR_{\quad\epsilon\lambda}^{\beta\quad\quad\rho}\widebar{\nabla}_\rho
	-\delta^{\alpha}_{\gamma}
	\bR_{\quad\lambda\epsilon}^{\rho\beta}\widebar{\nabla}_\rho
	+\bR_{\quad\lambda\epsilon}^{\beta\alpha}\widebar{\nabla}_\gamma
	\right\}+\nonumber\\
	&+2g_{5}\left\{-\delta^{\alpha}_{\gamma}
	\bR_{\quad\epsilon\lambda}^{\rho\quad\quad\beta}\widebar{\nabla}_\rho
	+\bR_{\quad\epsilon\lambda}^{\beta\quad\quad\alpha}\widebar{\nabla}_\gamma
	-\delta^{\alpha}_{\gamma}
	\bR_{\quad\lambda\epsilon}^{\rho\beta}\widebar{\nabla}_\rho
	+\delta^{\alpha}_{\gamma}
	\bR_{\quad\lambda\epsilon}^{\beta\rho}\widebar{\nabla}_\rho
	\right\}+\nonumber\\
	&+2g_{6}\left\{-\delta^\alpha_\gamma
	\widebar{R}^{+\beta}_\epsilon\widebar{\nabla}_\lambda
	+\delta^\alpha_\gamma\delta^\beta_\lambda
	\widebar{R}^{+\rho}_\epsilon \widebar{\nabla}_\rho-\delta^\alpha_\gamma
	\widebar{R}^{\beta}_{+\epsilon}\widebar{\nabla}_\lambda
	+\delta^\alpha_\lambda
	\widebar{R}^{\beta}_{+\epsilon}\widebar{\nabla}_\gamma \right\}+\nonumber\\
	&+2g_{7}\left\{-\delta^\alpha_\gamma
	\widebar{R}^{+\beta}_\epsilon\widebar{\nabla}_\lambda
	+\delta^\alpha_\lambda
	\widebar{R}^{+\beta}_{\epsilon}\widebar{\nabla}_\gamma-\delta^\alpha_\gamma
	\widebar{R}^{\beta}_{+\epsilon}\widebar{\nabla}_\lambda
	+\delta^\alpha_\gamma\delta^\beta_\lambda
	\widebar{R}^{\rho}_{+\epsilon} \widebar{\nabla}_\rho\right\}+\nonumber\\
	&+2g_{8}\left\{\bg_{\lambda\gamma}
	\bR_\epsilon^{-\beta}\widebar{\nabla}^\alpha
	-\bg_{\lambda\gamma}\bg^{\alpha\beta}
	\bR_\epsilon^{-\rho}\widebar{\nabla}_\rho-
	\bR_{\lambda\gamma\epsilon}^{\quad\quad\beta}\widebar{\nabla}^\alpha+\bg^{\alpha\beta}
	\bR_{\lambda\gamma\epsilon}^{\quad\quad\rho}\widebar{\nabla}_\rho
	\right.+\nonumber\\
	&\left.-\delta^\alpha_\gamma 
	\bR_\lambda^{-\beta}\widebar{\nabla}_\epsilon+\delta^\alpha_\gamma\delta^\beta_\epsilon
	\widebar{R}^{-\rho}_\lambda\widebar{\nabla}_\rho
	-\delta^\alpha_\gamma\bR_{\lambda\epsilon}^{-}\widebar{\nabla}^\beta
	+\bg^{\alpha\beta}\bR_{\lambda\epsilon}^{-}\widebar{\nabla}_\gamma\right\}+\nonumber\\
	&+2g_{9}\left\{-
	\bR_{\quad\gamma\epsilon\lambda}^{\beta}\widebar{\nabla}^\alpha
	+\bg^{\alpha\beta}
	\bR_{\quad\gamma\epsilon\lambda}^{\rho} \widebar{\nabla}_\rho-\delta^\alpha_\gamma
	\bR_{-\lambda}^{\beta}\widebar{\nabla}_\epsilon+\delta^\alpha_\gamma\delta^\beta_\epsilon
	\widebar{R}^{\rho}_{-\lambda}\widebar{\nabla}_\rho\right\}+\nonumber\\
	&+g_{10}\left\{-\delta^\alpha_\gamma\bR^{+\beta}_{\lambda}\widebar{\nabla}_\epsilon
	+\delta^\alpha_\gamma\delta^\beta_\epsilon
	\widebar{R}^{+\rho}_\lambda \widebar{\nabla}_\rho-
	\bR_{\quad\gamma\epsilon}^{\alpha\quad\quad\beta}\widebar{\nabla}_\lambda
	+\delta^\alpha_\l\bR_{\quad\gamma\epsilon}^{\beta\quad\quad\rho}\widebar{\nabla}_\rho
	\right.+\nonumber\\
	&\left.-\delta^\alpha_\gamma\bR^+_{\lambda\epsilon}\widebar{\nabla}^\beta
	+\bg^{\alpha\beta}\bR_{\lambda\epsilon}^{+}\widebar{\nabla}_\gamma
	-\delta^\alpha_\gamma\bR^\beta_{-\epsilon}\widebar{\nabla}_\lambda
	+\delta^\alpha_\lambda\bR^\beta_{-\epsilon}\widebar{\nabla}_\gamma\right\}+\nonumber\\
	&+g_{11}\left\{-\delta^\alpha_\gamma
	\bR^{\beta}_{+\lambda}\widebar{\nabla}_\epsilon
	+\delta^\alpha_\gamma\delta^\beta_\epsilon
	\widebar{R}^{\rho}_{+\lambda}\widebar{\nabla}_\rho-
	\bR_{\quad\gamma\epsilon}^{\alpha\quad\quad\beta}\widebar{\nabla}_\lambda
	+\delta^\alpha_\gamma\bR_{\quad\gamma\epsilon}^{\rho\quad\quad\beta}\widebar{\nabla}_\rho
	\right.+\nonumber\\
	&\left.-\delta^\alpha_\gamma\bR^+_{\epsilon\lambda}\widebar{\nabla}^\beta
	+\bg^{\alpha\beta}\bR_{\epsilon\lambda}^{+}\widebar{\nabla}_\gamma
	-\delta^\alpha_\gamma\bR^\beta_{-\epsilon}\widebar{\nabla}_\lambda
	+\delta^\alpha_\gamma\delta^\beta_\lambda\bR^\rho_{-\epsilon}\widebar{\nabla}_\rho\right\}+\nonumber\\
	&+2g_{12}\left\{-\bg^{\a\b}
	\widebar{R}^{+}_{\g\e}\widebar{\nabla}_\l+\d^\a_\l
	\widebar{R}^{+}_{\g\e}\widebar{\nabla}^\b-\d^\a_\g\d^\b_\e
	\widebar{R}\widebar{\nabla}_\l+\d^\a_\g\d^\b_\l \widebar{R}\widebar{\nabla}_\e\right\}\Bigg\}+\nonumber\\
	&+\{\a\leftrightarrow\b\}+\{\g\leftrightarrow\e\}
\eea
	The quadratic term involving gauge fluctuations with themselves reads
\bea
 K_{\l~~\t}^{\a\b\g\e}&=2g_{1}\left\{2\delta^\alpha_\tau\left(\bR_\lambda^{\gamma\beta\epsilon}-\bR_\lambda^{\quad\gamma\epsilon\beta}
\right)+2\left(\widebar{\nabla}^\e\widebar{\nabla}^{\b}\bg_{\l\tau}\bg^{\a\g}
-\widebar{\nabla}_\r\widebar{\nabla}^{\r}\bg_{\l\tau}\bg^{\a\g}\bg^{\b\e}
\right)\right\}+\nonumber\\
&+2g_{2}\left\{2\delta^\alpha_\tau\left(\bR_\lambda^{\quad\beta\gamma\epsilon}-\bR_\lambda^{\quad\epsilon\gamma\beta}
\right)+\left(\widebar{\nabla}^\e\widebar{\nabla}^{\b}\bg_{\l\tau}\bg^{\a\g}
-\widebar{\nabla}_\r\widebar{\nabla}^{\r}\bg_{\l\tau}\bg^{\a\g}\bg^{\b\e}
\right)\right\}+\nonumber\\
&+2g_{3}\left\{2\delta^\alpha_\tau\left(\bR^{\gamma\quad\beta\epsilon}_{\quad\lambda}
-\bR^{\gamma\quad\epsilon\beta}_{\quad\lambda}
\right)+2\left(\widebar{\nabla}^\e\widebar{\nabla}^{\b}\d^{\g}_{\l}\d^{\a}_{\t}
-\widebar{\nabla}_\r\widebar{\nabla}^{\r}\d^{\g}_{\l}\d^{\a}_{\t}\bg^{\b\e}
\right)\right\}+\nonumber\\
&+2g_{4}\left\{2\delta^\alpha_\tau\left(\bR^{\beta\gamma\quad\epsilon}_{\quad\lambda}
-\bR^{\epsilon\gamma\quad\beta}_{\quad\lambda}
\right)+\right.\nonumber\\
&\left.+\left(\widebar{\nabla}_\t\widebar{\nabla}^{\b}\d^{\e}_{\l}\bg^{\a\g}
-\widebar{\nabla}_\t\widebar{\nabla}_{\l}\bg^{\a\g}\bg^{\b\e}
+\widebar{\nabla}^\e\widebar{\nabla}_{\l}\d^{\b}_{\t}\bg^{\a\g}
-\widebar{\nabla}_\s\widebar{\nabla}^{\s}\d^{\e}_{\l}\d^{\b}_{\t}\bg^{\a\g}\right)\right\}+\nonumber\\
&+2g_{5}\left\{2\delta^\alpha_\tau\left(\bR^{\beta\epsilon\quad\gamma}_{\quad\lambda}
-\bR^{\epsilon\beta\quad\gamma}_{\quad\lambda}
\right)+\right.\nonumber\\
&\left.+\left(\widebar{\nabla}_\t\widebar{\nabla}^{\b}\d^{\e}_{\l}\bg^{\a\g}
-\widebar{\nabla}_\l\widebar{\nabla}_{\t}\bg^{\a\g}\bg^{\b\e}
+\widebar{\nabla}^\e\widebar{\nabla}_{\l}\d^{\b}_{\t}\bg^{\a\g}
-\widebar{\nabla}^\b\widebar{\nabla}^{\e}\d^{\g}_{\l}\d^{\a}_{\t}\right)\right\}+\nonumber\\
&+2g_{6}\left\{2\delta^\alpha_\tau\left(\delta^\beta_\lambda\bR^{\gamma\epsilon}_{+}
-\delta^\epsilon_\lambda\bR^{\gamma\beta}_{+}
\right)+\right.\nonumber\\
&\left.+\left(\widebar{\nabla}_\l\widebar{\nabla}^{\b}\d^{\e}_{\t}\bg^{\a\g}
-\widebar{\nabla}_\l\widebar{\nabla}_{\t}\bg^{\a\g}\bg^{\b\e}
+\widebar{\nabla}^\e\widebar{\nabla}_{\t}\d^{\b}_{\l}\bg^{\a\g}
-\widebar{\nabla}_\s\widebar{\nabla}^{\s}\d^{\b}_{\l}\d^{\e}_{\t}\bg^{\a\g}\right)\right\}+\nonumber\\
&+2g_{7}\left\{2\delta^\alpha_\tau\left(\delta^\beta_\lambda\bR^{\epsilon\gamma}_{+}
-\delta^\epsilon_\lambda\bR^{\beta\gamma}_{+}
\right)+\right.\nonumber\\
&\left.+\left(\widebar{\nabla}_\l\widebar{\nabla}^{\b}\d^{\e}_{\t}\bg^{\a\g}
-\widebar{\nabla}_\l\widebar{\nabla}_{\t}\bg^{\a\g}\bg^{\b\e}
+\widebar{\nabla}^\e\widebar{\nabla}_{\t}\d^{\b}_{\l}\bg^{\a\g}
-\widebar{\nabla}^\b\widebar{\nabla}^{\e}\d^{\a}_{\l}\d^{\g}_{\t}\right)\right\}+\nonumber\\
&+2g_{8}\left\{2\delta^\alpha_\tau\left(\bg^{\gamma\beta}\bR_\lambda^{-\epsilon}
-\bg^{\gamma\epsilon}\bR_\lambda^{-\beta}
\right)+\right.\nonumber\\
&\left.+\left(\widebar{\nabla}^\a\widebar{\nabla}^{\b}\bg_{\l\t}\bg^{\g\e}
-\widebar{\nabla}^\b\widebar{\nabla}^{\e}\bg_{\l\t}\bg^{\a\g}
+\widebar{\nabla}^\g\widebar{\nabla}^{\e}\bg_{\l\t}\bg^{\a\b}
-\widebar{\nabla}_\s\widebar{\nabla}^{\s}\bg_{\l\t}\bg^{\a\b}\bg^{\g\e}\right)\right\}+\nonumber\\
&+2g_{9}\left\{2\delta^\alpha_\tau\left(\bg^{\gamma\beta}\bR^\epsilon_{-\lambda}
-\bg^{\gamma\epsilon}\bR^{\beta}_{-\lambda}
\right)+\right.\nonumber\\
&\left.+\left(\widebar{\nabla}^\b\widebar{\nabla}_{\l}\d^{\a}_{\t}\bg^{\g\e}
-\widebar{\nabla}^\b\widebar{\nabla}^{\e}\d^{\g}_{\t}\d^{\a}_{\l}
+\widebar{\nabla}_\t\widebar{\nabla}^{\e}\d^{\g}_{\l}\bg^{\a\b}
-\widebar{\nabla}_\t\widebar{\nabla}_{\l}\bg^{\a\b}\bg^{\g\e}\right)\right\}+\nonumber\\
&+2g_{10}\left\{\delta^\alpha_\tau\left(\bg^{\gamma\beta}\bR^{+\epsilon}_{\lambda}
-\bg^{\gamma\epsilon}\bR^{+\beta}_{\lambda}
+\delta^\beta_\lambda\bR^{\gamma\epsilon}_{-}-\delta^\epsilon_\lambda\bR^{\gamma\beta}_{-}\right)+\right.\nonumber\\
&+\left.+\left(\widebar{\nabla}_\l\widebar{\nabla}^{\b}\d^{\a}_{\t}\bg^{\g\e}
-\widebar{\nabla}_\l\widebar{\nabla}^{\e}\d^{\a}_{\t}\bg^{\b\g}
+\widebar{\nabla}^\e\widebar{\nabla}^{\g}\d^{\b}_{\l}\d^{\a}_{\t}
-\widebar{\nabla}_\s\widebar{\nabla}^{\s}\d^{\b}_{\l}\d^{\a}_{\t}\bg^{\g\e}\right)\right\}+\nonumber\\
&+2g_{11}\left\{\delta^\alpha_\tau\left(\bg^{\gamma\beta}\bR^\epsilon_{+\lambda}
-\bg^{\gamma\epsilon}\bR^\beta_{+\lambda}
+\delta^\beta_\lambda\bR^{\epsilon\gamma}_{-}-\delta^\epsilon_\lambda\bR^{\beta\gamma}_{-}\right)+\right.\nonumber\\
&\left.+\left(\widebar{\nabla}_\l\widebar{\nabla}^{\b}\d^{\a}_{\t}\bg^{\g\e}
-\widebar{\nabla}_\l\widebar{\nabla}^{\e}\d^{\a}_{\t}\bg^{\b\g}
+\widebar{\nabla}_\t\widebar{\nabla}^{\e}\d^{\b}_{\l}\bg^{\a\g}
-\widebar{\nabla}_\t\widebar{\nabla}^{\b}\d^{\a}_{\l}\bg^{\g\e}\right)\right\}+\nonumber\\
&+2g_{12}\left\{2\delta^\alpha_\tau\left(\delta^\beta_\lambda\bg^{\gamma\epsilon}\bR
-\delta^\epsilon_\lambda\bg^{\gamma\beta}\bR
\right)+\right.\nonumber\\
&\left.+\left(\widebar{\nabla}_\l\widebar{\nabla}^{\g}\d^{\e}_{\t}\bg^{\a\b}
-\widebar{\nabla}_\l\widebar{\nabla}_{\t}\bg^{\a\b}\bg^{\g\e}
+\widebar{\nabla}^\a\widebar{\nabla}_{\t}\d^{\b}_{\l}\bg^{\g\e}
-\widebar{\nabla}^\a\widebar{\nabla}^{\g}\d^{\b}_{\l}\d^{\e}_{\t}\right)\right\}+\{\a\leftrightarrow\b,\g\leftrightarrow\e,\l\a\b\leftrightarrow\t\g\e\}
\eea
\section{Metric from connection.}

The problem of determining the metric structure of the space-time manifold out of free-falling observations has been in the forefront of research at least since the pioneering work of Weyl, on the mathematical side, and Ehlers, Pirani and Schild, on the physics side {\em confer} \cite{Ehlers} and references therein.
\par

There are several aspects.
First of all, the connection (without any use of the metric) determines uniquely the parallel propagator along a given curve.
To be specific, this is given by a Wilson line, a path ordered exponential
\be
 g^\a_{~\beta^\prime}(x,\xp)\equiv P~\left[e^{\scaleint{4.5ex}_i^f~\Gamma_\m\dot{x}^\m~d\t}\right]^\a_{~\beta^\prime}
\ee
where 
\be
\Gamma_\m\equiv \left(\Gamma_\m\right)^\a_{\b^\prime}
\ee
and the integral is done  through a curve
\be
x^\m=x^\m(\t)
\ee
where
\begin{align}
&x^\m(\t_i) =x^\m\\
&x^\m(\t_f)=x^{\mup}
\end{align}

\par

Nevertheless, not every connection is metric-compatible; that is, it is not always possible to find a metric such that the given connection (even assumed to be  torsion-free) is the Levi-Civita one stemming from the metric itself.

The condition for that to be true can be  clearly stated using the  Christoffel's symbols of
first kind, namely 
\be 
\pd_\m\bigg(\left\{\d;\b
\l\right\}+\left\{\b;\l \d\right\}\bigg)=\pd_\l\bigg(\left\{\d;\b
\m \right\}+\left\{\b;\d \m\right\}\bigg) 
\ee
 which expresses the
obvious fact that 
\be \pd_\m\pd_\l~ g_{\d\b}=\pd_\l\pd_\m
~g_{\b\d} \ee In order to determine the generated metric in such
cases as it exists, (that is, when the integrability condition is
fulfilled), there is the linear  system  of partial differential equations 
\be 
\pd_\l~g_{\d\b}=g_{\a\d}~ \Gamma^\a_{\b\l}+ g_{\a\b}~\Gamma^\a_{\l\d} 
\ee
whose trace implies
\be
g^{\d\b}~\pd_\l~ g_{\d\b}= 2 ~\Gamma^\b_{\b\l}
\ee
The integrability conditions for such a system are precisely as
above, namely

\be 
\pd_\m\bigg(g_{\d\a}\Gamma^\a_{\b \l}+g_{\b\a}\Gamma^\a_{\l
\d}\bigg)=\pd_\l\bigg(g_{\d\a}\Gamma^\a_{\b \m}
+g_{\a\b}\Gamma^\a_{\d \m}\bigg) 
\ee 
 At the
linearized level, assuming 
\bea
&g_{\a\b}\equiv \eta_{\a\b}+\kappa h_{\a\b}\nonumber\\
&\Gamma_{\a\b\g}=O(\kappa) 
\eea
 The integrability condition reads
  \be
\pd_\m\left(\Gamma_{\d\b\l}+\Gamma_{\b\d\l}\right)=\pd_\l\left(\Gamma_{\d\b\m}+\Gamma_{\b\m\d}\right)
\ee
 This can be written in a suggestive way as 
 \be
\pd_\m\Gamma_{\d\b\l}-\pd_\l \Gamma_{\d\b\m}=\pd_\l \Gamma_{\b\d\m}-\pd_\m
\Gamma_{\b\d\l} 
\ee 
or introducing the one-forms 
\be 
\chi_{\a\b}\equiv
\Gamma_{\a\b\l} dx^\l 
\ee 
this is equivalent to  a certain one-form to
be closed, that is, 
\be 
d~\chi_{(\a\b)}=0 
\ee 
It is always
possible to write the connection as 
\bea
&\Gamma_{\a\b\l}\equiv{1\over
4}\left(\Gamma^+_{\a\b\l}+\Gamma^-_{\a\b\l}\right) \eea where \bea
&\Gamma^+_{\a\b\l}\equiv\Gamma_{\a\b\l}+\Gamma_{\b\a\l}+\Gamma_{\a\l\b}+\Gamma_{\b\l\a}\nonumber\\
&\Gamma^-_{\a\b\l}\equiv
\Gamma_{\a\b\l}+\Gamma_{\a\l\b}-\Gamma_{\b\a\l}-\Gamma_{\b\l\a}
 \eea
  The preceding
identity then implies
 \be 
 \Gamma^+_{\a\b\l}=~\pd_\l \phi_{\a\b}
  \ee
Once this condition is fulfilled, the solution is given by the
solution of the first order linear differential equation
 \be
\pd_\l h_{\d\b}=\Gamma_{\d\b\l}+\Gamma_{\b\d\l}={1\over 2}~\pd_\l
\phi_{\d\b} 
\ee
 It follows that 
 \be h_{\d\b}={1\over
2}~\phi_{\d\b}+ C 
\ee 
This also shows that
 \be 
 \Gamma^-_{\a\b\l}
  \ee
is pure gauge, and the physical metric is independent of it.

\section{Constant curvature spaces.}\label{ccs}
For constant curvature spaces the n-dimensional Riemann tensor obeys

\be 
R_{\m\n\r\s}=-{2\l\over
(n-1)(n-2)}\left(g_{\m\r}~g_{\n\s}-g_{\m\s}~g_{\n\r}\right)\equiv \pm
{1\over L^2}~\left(g_{\m\r}~g_{\n\s}-g_{\m\s}~g_{\n\r}\right)
 \ee
where $x^\m,\; \m=1,\ldots n$. It is useful to work with the
 Synge's \cite{Synge} world function $\Omega(x,y)$ which is defined as
 \be
\Omega (x,\xp)={\frac {1}{2}}\int_0^1 g_{\mu \nu }(z)t^{\mu
}t^{\nu }d\lambda  \equiv{1\over 2}~s_{x,\xp}^2 
\ee 
where $x$ and
$\xp$ are two points close enough so that there is a unique geodesic
joining them $\g$, parametrized by an affine parameter $\l$ such
that
 \bea
&\g(0)=x\nonumber\\
&\g(1)=\xp 
\eea
 The only advantage of the world function over the
arc is that the former is always real (although sometimes negative) even
in pseudoriemannian spaces. This is not an issue on Riemannian spaces (like the sphere)
though, in which case is actually simpler to work with the arc length, $s$.
\par
The basic equation that determines the world function in general
is 
\be 
g^{\m\n}~\pd_\m \Omega ~\pd_\n \Omega\equiv \Omega^\m~\Omega_\m=2~\Omega 
\ee
this just because 
\be 
\Omega={1\over 2} s^2\quad \Rightarrow\quad
\Omega_\m=s s_\m\quad \Rightarrow g^{\m\n} \Omega_\m \Omega_\n=s^2=2\Omega
\ee
\par
It follows that
 \be\label{uno} 
 \Omega_{\m\n}~\Omega^\m\equiv
\nabla_\n\nabla_\m\Omega ~\nabla^\m \Omega=\Omega_\n 
\ee 
It is also the case
that 
\be\label{dos}
 \left[\nabla_\l,\nabla_\n\right]~s_\m =
R_{\l\n\m\r}s^\r=\pm{1\over
L^2}~\left(g_{\l\m} s_\n- s_\l~g_{\n\m}\right) 
\ee
(where $s_\m\equiv {\Omega_\m\over s}$). Please note that in this equation all indices are covariant ones.
\par
 Invariant
tensors can be expanded in outer products of $s_\m$ (or $\Omega_\m$) and
$g_{\m\n}$, with coefficients that depend on $s$ only. For example \cite{Allen}
\be
s_{\m\n}={1\over L\tan~{s\over L}}\left(g_{\m\n}-s_\m s_\n\right)
\ee
which implies
\be
\Box s={n-1\over L\tan~{s\over L}}
\ee
Let us work explicitly a couple of examples (in the case of the n-sphere $S_n$, to be specific).
\paragraph{Example 1} The inverse of the d'Alembertian, $G\equiv \Box\,^{-1}$
\be
\Box G(s)=\d^n(x)
\ee
The ODE to be solved is
\be
G^{\prime\prime}(s)+{n-1\over L\tan{s\over L}}~G^\prime(s)=\d^n(x)
\ee
When $s\neq 0$
\be
L~\tan~{s\over L}~G^{\prime\prime}(s)+(n-1)~G^\prime(s)=0
\ee
When $s\sim 0$
\be
G(s)\sim s^{2-n}
\ee
which is the correct behavior for a Dirac delta singularity.
\par
The exact solution reads
\be
G(s)=C_1+C_2 \cos~{s\over L }~\,_2F_1\left({1\over 2},{n\over 2};{3\over 2},\cos^2~{s\over L}\right)\label{vanilla1}
\ee

\paragraph{Example 2}Let us now compute a vector Green's function. In order to do that, it is best to first compute the inverse of the second power of the d'Alembertian, $G_2\equiv \Box\,^{-2}$. Let us start with
\be
\left(\Box\d_\m^\n+a\nabla_\m\nabla^\n\right)~G_\n^\s(s)=\d_\m^\s ~\d^n(x)\label{vanilla2}
\ee
Please notice that on the sphere there is no zero mode even for $a=-1$, because
\be
\nabla^\m \Box-\nabla_\l\nabla^\m\nabla^\l=-{1\over L^2}~(n-1)\nabla^\m
\ee
We shall need

\be
\Box^2 G_2(s)=\d^n(x)
\ee

which is equivalent to

\be
G_2^{IV}(s)+\frac{2(n-1)}{L\tan~{s\over L}}~G_2'''(s)+\frac{(n-1)((n-1)\cos^2~{s\over L}-2)}{L^2 \sin^2~{s\over L}}~G_2''(s)+\frac{(n-1)(3-n)\cos~{s\over L}}{L^3\sin^3~{s\over L}}~G_2'(s)=0\ee

 This can be rewritten as

\be \Box (\Box G_2(s))=\left(\dfrac{d^2}{ds^2}+{n-1\over L\tan{s\over L}}~\dfrac{d}{ds}\right)\left(\dfrac{d^2}{ds^2}G_2(s)+{n-1\over L\tan{s\over L}}~\dfrac{d}{ds}G_2(s)\right)=\d^n(x)\ee

And using the result \eqref{vanilla1} we get that the general solution must have the form

\be
G_2(s)=G(s)+h(s)
\ee

Where $h(s)$ is the solution of the equation:

\be
\left(\dfrac{d^2}{ds^2}h(s)+{n-1\over L\tan{s\over L}}~\dfrac{d}{ds}h(s)\right)=G(s)
\ee

We can illustrate how to get the solution of \eqref{vanilla2} by studying the case  when $L\to\infty$ (flat space). In that case the general solution is given by

\be
G_2(s)=c_1\frac{s^{2-n}}{2-n}+c_2\frac{s^{4-n}}{4-n}+c_3\frac{s^2}{2}+c_4\ee

Then, for $\l=-\dfrac{a}{1+a}$ we can obtain our solution

\bea
G^{\n}_\m(s)&=\left(\Box\d_\m^\n+\l\partial_\m\partial^\n\right)~G_2(s)=-\l\left[c_1n s^{-n}+c_2(n-2) s^{2-n}\right]s_\m s^\n+\nonumber\\
&+\left[\l c_1 s^{-n}+c_2(2+\l)
s^{2-n}+(n+\l)c_3\right]\d_{\m}^{\n}\nonumber\\\eea

Finally, in the case of $n=4$, we need $c_1=0$ to recover the correct behaviour when $s\to 0$, and since we can set $c_3=0$ as it enters as an additive constant we get the desired tensor Green's function as
\bea
G^{\n}_\m(s)&=c_2\frac{2+a}{1+a}\left(\frac{1}{s^2}\d^\n_\m+{2a\over
	a+2}~\frac{s^\n s_{\m}}{s^2}\right)\eea



\end{document}